\documentclass[preprint,amsmath,amssymb,aps,a4]{revtex4}
\usepackage{graphicx}
\usepackage{bm}
\usepackage{txfonts}
\usepackage{hyperref}
\usepackage{calligra}
\usepackage{amsmath}
\usepackage{mathrsfs}
\usepackage{amssymb}
\usepackage[english]{babel}

\date{\today}
\newcommand{\be}{\begin{eqnarray}}
	\newcommand{\ee}{\end{eqnarray}}

\newcommand{\bfz}{{\bf 0}_{\perp}}
\newcommand{\bfa}{{\bf a}_{\perp}}

\newcommand{\bfk}{{\bf k}_{\perp}}

\newcommand{\bfki}{{\bf k}_{\perp i}}
\newcommand{\bfb}{{\bf b}_{\perp}}

\newcommand{\bfP}{{\bf P}_{\perp}}

\newcommand{\bfpi}{{\bf p}_{\perp i}}
\newcommand{\bfpip}{{\bf p}_{\perp i}^\prime}

\newcommand{\Dp}{{\bf \Delta}_{\perp}}

\newcommand{\lipr}{{\lambda_{i}}^\prime}

\usepackage{xcolor}
\begin{document}
	
\title{Generalized parton distributions for low-lying octet baryons with non-zero skewness}
	
\author{Navpreet Kaur}
\email{knavpreet.hep@gmail.com}
\affiliation{Department of Physics, Dr. B.R. Ambedkar National
Institute of Technology, Jalandhar, 144008, India}

\author{Harleen Dahiya}
\email{dahiyah@nitj.ac.in}
\affiliation{Department of Physics, Dr. B.R. Ambedkar National
Institute of Technology, Jalandhar, 144008, India}
	
\date{\today}%
\begin{abstract}
We have performed the structural analysis of low-lying octet baryon members by assuming charge-isospin symmetry among quark flavors in transverse and longitudinal space. Using the light-cone spectator diquark  model, we have investigated the electromagnetic form factors of the octet baryons and have compared them with the available data. Model scale has been estimated by comparing the valence $u$ quark's parton distribution function (PDF) of proton with the result of CTEQ5L parameterization. A comparative analysis of PDFs and generalized parton distributions (GPDs) for non-zero skewness among octet baryons has also been demonstrated. Fourier transform of the deeply virtual Compton scattering amplitude with respect to skewness has been further applied to study the diffraction patterns in the boost-invariant longitudinal position space.
\end{abstract}
%
\maketitle
%
%
\section{Introduction\label{secintro}}

The study of hadron structure,  with the interactions among partons (that include quarks $q$ and gluons $g$) inside these hadrons, make their structure complex in quantum chromodynamics and acts as a big hurdle for researchers to understand their static and dynamic properties. This complex internal structure of hadrons can be scrutinized with the help of generalized parton correlation function (GPCF) which is the fully unintegrated and off-diagonal quark-quark correlator \cite{Lorce:2013pza}. Integration of GPCF over the light-cone component of quark's momentum leads to the generalized transverse momentum-dependent distributions (GTMDs) that further provide a multi-dimensional picture of a hadronic system \cite{Echevarria:2016mrc}. This complex internal structure of hadrons can be viewed three-dimensionally via partonic transverse-momentum dependent distributions (TMDs) and generalized parton distribution (GPDs) \cite{Meissner:2009ww}. These distributions can respectively be derived from the GTMDs by assuming either zero momentum transfer or by integrating over the transverse momentum $\bfk$ of partons. TMDs are the unintegrated parton distribution functions (PDFs) \cite{Collins:2007ph} that portray the probability of finding a parton inside a hadron with longitudinal momentum fraction $x$ and transverse momentum $\bfk$ with respect to the direction of its parent hadron momentum \cite{Collins:1981uw}. The past few years have witnessed considerable progress in both theoretical and experimental aspects of TMDs \cite{Cerutti:2022lmb, Bacchetta:2024fci, Ji:2004wu, HERMES:2003gbu, SpinMuon:1997yns, Puhan:2023ekt, Sharma:2023wha, Sharma:2022ylk, Bacchetta:2019sam}. On the other hand, the work of Ji, Collins, Radyushkin, and Blumlein showcases the cutting-edge developments during the early stage of the GPDs \cite{Ji:1998pc, Braunschweig:1985nr, Collins:1996fb, Radyushkin:1996ru, Blumlein:1999sc}. GPDs can depict the three-dimensional structure of partons in momentum space as a function of three variables: longitudinal momentum fraction $x$, skewness parameter $\zeta$ and invariant moment transfer $t$. Here, the skewness parameter reflects the longitudinal momentum transfer. 
\par 

GPDs are non-forward matrix elements that are accessed experimentally via deeply virtual Compton scattering (DVCS) \cite{Hessberger:2016ypd, H1:2007vrx, JeffersonLabHallA:2006prd, Collins:1998be} and deeply virtual meson production (DVMP) \cite{Goloskokov:2007nt, Vanderhaeghen:1999xj} processes. Several experiments, conducted by H$1$ collaboration \cite{H1:2001nez}, ZEUS collaboration \cite{ZEUS:2003pwh} and fixed target experiment at HERMES \cite{HERMES:2012gbh, HERMES:2011bou} have completed DVCS data collection. DVCS exhibits similarities to the classic Compton scattering process with a key distinction of involving highly virtual (off-shell) incoming photon to scatter off a hadron target with a real (on-shell) outgoing photon. Therefore, the skewness can not be zero in physical experiments and an analysis of the skewness parameter is crucial. In the forward limit of zero momentum transfer and zero skewness, GPDs are reduced to ordinary PDFs which are functions of one-dimensional longitudinal momentum fraction $x$ only. Whereas, electromagnetic form factors (EMFFs) of the local current can be obtained by taking the first moment of GPDs in $x$ with zero skewness.
\par 
The distributions of GPDs can be defined in the interval $x \in [-1,1]$ and categorized into  Efremov-Radyushkin-Brodsky-Lepage (ERBL) and Dokshitzer-Gribov-Lipatov-Altarelli-Parisi (DGLAP) regions \cite{Diehl:2003ny}. DGLAP region has a domain of $x \in [-1,-\zeta]$ and $x \in [\zeta,1]$, whereas ERBL has a domain of $x \in [-\zeta,\zeta]$.
Since the region $-1 < x < -\zeta$ corresponds to the interaction of an anti-quark and $0 < x < \zeta$ includes the particle number changing interactions respectively, we have chosen to work within this DGLAP region ($x \in [\zeta,1]$). This includes the emission and re-absorption of a quark (gluon) after the  interaction with a highly virtual photon. With different approaches, evolved distributions have been studied to reconcile with experimental data at high $Q^2$ \cite{Mamo:2022jhp, Kumericki:2007sa, Mueller:2005ed, Freund:2001bf}. Along with this, considerable attention has also been directed towards the higher twist GPDs \cite{Duplancic:2023xrt, Sharma:2023ibp, Pire:2013vea}. In the longitudinal position space, DVCS amplitudes can be studied by Fourier transforming (FT) $\zeta$ to $\sigma$. Physically, these FT amplitudes of GPDs measure the correlation between outgoing and incoming quark currents at fixed light-cone time $\tau$ with transverse separation $\bfb$ and longitudinal separation $\sigma=b^- P^+ /2$ \cite{Kaur:2018ewq, Mondal:2017wbf, Kumar:2015yta, Manohar:2010zm, Chakrabarti:2008mw, Brodsky:2006ku, Brodsky:2006in}. It exhibits diffraction patterns analogous to the diffractive scattering of a wave in optics that relates the physical size of the scattering center in one-dimension. 
\par 
Theoretically as well as experimentally, an enormous amount of work has been done in the past and is presently going on to scrutinize the complex internal structure of the nucleons \cite{Barry:2023qqh, Bor:2022fga, GlueX:2019mkq, Moutarde:2019tqa, HERMES:2001bob} which are the members of baryon octet and carry spin-parity quantum number as $J^P=\big(\frac{1}{2}\big)^+$. Nucleons are assumed to be composed of three light quarks whereas, other members of the baryon octet are considered to constitute strange quark(s) along with $u$ and $d$ quarks. Due to a shorter life span of strange baryons, their experimental data related to DVCS or DVMP processes are not yet available. In the recent years, few physicists have taken the initiative to explore the internal structure of other members of low-lying baryon octet using lattice simulations \cite{CSSM:2014knt, Shanahan:2014cga} and phenomenological models \cite{Han:2024ucv, Zhu:2023nhl, Carrillo-Serrano:2016igi, Zhang:2016qqg, Jiang:2009jn}. The analysis of strange baryon dynamics can provide valuable insights into the hypernuclear physics \cite{Feliciello:2015dua} and is also crucial to study the properties of neutron stars and strange stars \cite{Wang:2005vg}. Taking motivation from this growing domain of particle physics, we have studied the EMFFs, PDFs, non-zero skewed GPDs, and their distributions in longitudinal position space. 
\par  
 In order to study a complex internal structure of strange baryons relativistically, we have employed the light-cone formalism \cite{Dirac:1949cp}. Instead of a three-body system, we treat a baryon being composed of an active quark and a spectator diquark. This active quark will interact with an incoming virtual photon. We have chosen a light-cone diquark spectator model as it has been successfully  employed to study the TMDs \cite{Bacchetta:2008af}, fragmentation functions \cite{Jakob:1997wg} and Sivers functions \cite{Bacchetta:2003rz} with different summations over polarization vectors of diquarks. For realistic flavor analysis, we have considered the axial vector diquarks as isoscalar and isovector spectator diquarks. Along with the transverse and longitudinal polarization of diquarks, we have also considered the timelike polarized diquarks to see their impact on the distributions. Assuming charge-isospin symmetry \cite{Zhang:2016qqg, Jakob:1997wg}, we have considered the probabilistic weights $3:1:2$ among the scalar isoscalar, vector isoscalar and vector isovector configurations. 
\par
The present paper is organized as follows. The light-cone formalism and the diquark spectator model have been detailed in Sec. \ref{SecModel}. Numerical parameters used in the present  calculations have been presented in Sec. \ref{secnumpar}. Evaluation of EFFs and their comparison with lattice data have been shown in Sec. \ref{SecEMFFs}. In Sec. \ref{SecGPDs}, we have expressed the explicit expressions of chiral even GPDs. For the case of proton $p$, the PDF of $u$ quark flavor is compared with the result of CTEQ5L parameterization in order to estimate the model scale. A discussion of comparative analysis among other octet baryon members has also been presented in the same section. Sec. \ref{SecLong_space} discusses the GPDs in longitudinal position space. Summary of the work has been presented in Sec. \ref{SecSumm}.
\section{Model description\label{SecModel}} 
In this section, we provide the essential details of the adopted formalism to study the quark-diquark system of baryons $X$. For a baryon, moving with total momentum $P$, the expansion of its Fock eigenstate on the complete basis of the constituent eigenstates can be expressed as \cite{Brodsky:2000ii}
\be
|\mathcal{B}(P^{+},{\bf P_\perp^2)}\rangle &=& \sum_{\mathcal{N}} \prod_{i=1}^{\mathcal{N}} \frac{dx_i~  d^2\bfki}{2(2\pi)^3\sqrt{x_{i}}} \, 16 \pi^{3} \, \delta \bigg(1-\sum_{i=1}^{\mathcal{N}} x_{i}\bigg) \, \delta^{(2)} \bigg(\sum_{i=1}^{\mathcal{N}}\bfki\bigg) \nonumber \\		
&\times& \psi_{\mathcal{N}}(x_{i},\bfki,\lambda_{i})|\mathcal{N}; x_{i} P^{+},x_{i}\bfP+\bfki,\lambda_{i}\rangle \, ,
\ee
where $x_i=k_i^+/P^+$ and $\lambda_i$ respectively denote the light-cone longitudinal momentum fraction and helicity carried by \textit{i}th constituent of a baryon. The term $(x_{i}\bfP+\bfki)$ represents the physical transverse momenta $\bfpi$. The momentum of each constituent of baryon is given by $k_i$ whose light-cone coordinates are 
\begin{equation}
	k_{i}=(k_{i}^{+},k_{i}^{-},\bfki)=\bigg(x_{i}P^{+},\frac{\bfki^{2}+m_{i}^{2}}{x_{i}P^{+}},\bfki\bigg) \, .
\end{equation}
We also adopt light-cone gauge, $A^{+}=0$. The $\mathcal{N}$-particle Fock states are normalized as
\begin{equation}
	\langle \mathcal{N}; p_{i}^{\prime +}, \bfpip, \lambda_{i}^{\prime} |\mathcal{N};  p_{i}^{+},\bfpi,\lambda_{i}\rangle = \prod_{i=1}^{\mathcal{N}} 16 \pi^{3} \, p_{i}^{+} \, \delta(p_{i}^{\prime +}-p_{i}^{+}) \, \delta^{(2)} (\bfpip-\bfpi) \, {\delta}_{\lipr \lambda_{i}} \, .
\end{equation}
As the light-cone coordinates of total momentum, $P^+$ and $\bfP$ are kinematical in nature, the Fock projection of constituent on a baryon eigenstate gives the universal and process independent light-cone wave functions $\psi_{\mathcal{N}}^{J_z}$ with $J_z$ as the projection of a baryon. We have considered a frame in which light-cone four-vectors are defined as $a=[a^+,a^-,\bfa]$. We choose an asymmetric light-cone frame whose initial and final state coordinates of a baryon are respectively represented as
\be 
P &=& \bigg(P^+,\frac{M^2_X}{P^+},\bfz \bigg) \, , \nonumber \\
P^\prime &=& \bigg((1-\zeta)P^+,\frac{M^2_X+\Dp^2}{(1-\zeta)P^+},\Dp \bigg) \, ,
\ee 
where $\bar{P}=\frac{P^+ + P^{\prime +}}{2}$ represents the average four-vector momentum of a baryon and $\Delta=(P-P^\prime)$ represents the momentum transfer. The invariant momentum transfer is given by $t=t_0-\Dp^2/(1-\zeta)$ with $t_0=-\zeta^2 M^2_X/(1-\zeta)$, the skewness parameter is $\zeta=-\Delta^+/2P^+$ and $M_X$ denotes the mass of a baryon $X$. The instant form SU($6$) quark-diquark wave function for the members of a baryon octet can be written as \cite{Lichtenberg:1968zz}
\be 
|\mathcal{B}\rangle^{\Uparrow,\Downarrow} = \cos \theta \sum_q a_q |q_1 S(q_2 q_3) \rangle ^{\Uparrow,\Downarrow} + \sin \theta \sum_{q^\prime} b_q^\prime |q_1^\prime V(q_2^\prime q_3^\prime) \rangle ^{\Uparrow,\Downarrow} \, ,
\label{InstantWfn}
\ee 
where the summation is over different quark-diquark pairs. $|q_1 S(q_2 q_3) \rangle$ and $|q_1^\prime V(q_2^\prime q_3^\prime) \rangle$ represent the quark-scalar diquark and quark-vector diquark pairs respectively. $a_q$ and $b_q$ are the coefficients satisfying the normalization condition. The mixing angle attributed to the spin flavor SU($6$) symmetry breaking is represented by $\theta$. For the symmetric case, it is chosen as $\pi/4$. $\Uparrow,\Downarrow$ denote the spin projections $J_z=\pm \frac{1}{2}$ of a baryon, labeled as $\lambda_X$. For the case of proton $p$, the instant form of the  wave function is written as
\be 
|p\rangle^{\Uparrow,\Downarrow} = \frac{1}{\sqrt{2}} |u S(ud)\rangle^{\Uparrow,\Downarrow} -\frac{1}{\sqrt{6}} |u V(ud)\rangle^{\Uparrow,\Downarrow} +\frac{1}{\sqrt{3}} |d V(uu)\rangle ^{\Uparrow,\Downarrow} \, .
\ee
The instant form of the wave functions for  $\Sigma$ and $\Xi$ can similarly be written as
\be 
|\Sigma^+\rangle^{\Uparrow,\Downarrow} = \frac{1}{\sqrt{2}} |u S(us)\rangle^{\Uparrow,\Downarrow} -\frac{1}{\sqrt{6}} |u V(us)\rangle^{\Uparrow,\Downarrow} +\frac{1}{\sqrt{3}} |s V(uu)\rangle ^{\Uparrow,\Downarrow} \, ,  \\
|\Xi^+\rangle^{\Uparrow,\Downarrow} = \frac{1}{\sqrt{2}} |s S(us)\rangle^{\Uparrow,\Downarrow} +\frac{1}{\sqrt{6}} |s V(us)\rangle^{\Uparrow,\Downarrow} -\frac{1}{\sqrt{3}} |u V(ss)\rangle ^{\Uparrow,\Downarrow} \, .
\ee
Since the spin coupling will be introduced in the derivation of the wave function, the flavor decomposition for the case of proton can be expressed as \cite{Zhang:2016qqg, Jakob:1997wg}
\be 
f_1^{p_u} &=& \frac{3}{2} f_1^\mathfrak{s} + \frac{1}{2} f_1^\mathfrak{a} \, , \nonumber \\
f_1^{p_d} &=& f_1^\mathfrak{a} \,.
\ee 
These equations imply that for the valence quark distribution of $u$, both scalar and axial vector Fock states contribute, whereas for the $d$ quark valence distribution, only the vector Fock state contributes. Analogously, the wave functions of other members of the octet baryons can be formulated with the replacement of appropriate quark flavor content and coefficients following Ref. \cite{Zhang:2016qqg}. The two particle Fock state expansion for the case of scalar diquark is expressed as
\be 
|q S\rangle^{\Uparrow , \Downarrow}_X = \int \frac{dx d^2\bfk}{16 \pi^3 \sqrt{x(1-x)}} ~ \sum_{\lambda_q} \psi^{\Uparrow, \Downarrow X}_{\lambda_q} (x,\bfk) ~\big|\lambda_q,x P^+,\bfk \rangle  \, ,
\ee 
where $\lambda_q$ (= $\uparrow,\downarrow$) denotes the helicity of an active quark. In the similar vein, the two particle Fock state expansion for the case of axial vector diquark is expressed as
\be 
|q V\rangle^{\Uparrow , \Downarrow}_X = \int \frac{dx d^2\bfk}{16 \pi^3 \sqrt{x(1-x)}} \sum_{\lambda_q} \sum_{\lambda_a} \psi^{\Uparrow, \Downarrow X}_{\lambda_q \lambda_{\mathfrak{a}}} (x,\bfk) ~\big|\lambda_q \lambda_{\mathfrak{a}},x P^+,\bfk \rangle   \, ,
\ee 
with $\lambda_{\mathfrak{a}}$ $(= \pm,0,\mathcal{T})$ as the helicitiy of a diquark. Here, $\pm$, $0$ and $\mathcal{T}$ represent the transverse, longitudinal and timelike polarization respectively. In accordance with Ref. \cite{Bacchetta:2008af}, light-cone wave functions for scalar and axial vector diquarks are respectively defined as
\be 
\psi^{\lambda_X X}_{\lambda_q} (x,\bfk) = \sqrt{\frac{k^+}{(P-k)^+}} \frac{1}{k^2-m^2_q} \bar{u} (k,\lambda_q) \mathcal{Y}_{\mathfrak{s}} U(P,\lambda_X) \, ,
\label{ScalarWfn}
\ee
and 
\be 
\psi^{\lambda_X X}_{\lambda_q \lambda_{\mathfrak{a}}}  (x,\bfk) = \sqrt{\frac{k^+}{(P-k)^+}} \frac{1}{k^2-m^2_q} \bar{u} (k,\lambda_q) \epsilon^\ast_\mu (P-k,\lambda_{\mathfrak{a}}) \cdot \mathcal{Y}_{\mathfrak{a}}^\mu U(P,\lambda_X) \, .
\label{VectorWfn}
\ee 
Here, $u(k,\lambda_q)$ and $U(P,\lambda_X)$ represent the spin-$1/2$ Dirac spinors for a quark (with mass $m_q$) and baryon respectively with $k$ and $P$ being their respective total momenta. The term $\epsilon_\mu (P-k,\lambda_{\mathfrak{a}})$ is a four-vector polarization of spin-$1$ diquark with total momentum $(P-k)$. The expressions corresponding to $\epsilon_\mu (P-k,\lambda_{\mathfrak{a}})$ have been presented in the Appendix. To conserve the  ``spin sum rule", helicities must follow the constraints imposed by angular momentum conservation
\be 
\lambda_X=\lambda_q+\lambda_D+L_z\,,
\ee
where $D$ represents the diquark that can be scalar $\mathfrak{s}$ and axial vector $\mathfrak{a}$, whereas, $L_z$ corresponds to the relative orbital angular momentum projection between the quark and the diquark. The baryon-quark-diquark vertexes for scalar and axial vector diquark are respectively expressed as
\be 
\mathcal{Y}_s = i \, g_{X\mathfrak{s}} (k^2) \textbf{I} \, ,
\label{scalar}
\ee
and 
\be 
\mathcal{Y}_a^\mu = i \, \frac{g_{X_\mathfrak{a}} (k^2)}{\sqrt{2}} \gamma^\mu \gamma_5 \, .
\label{vector}
\ee
Here, $g_{X_D} (k^2)$  corresponds to the scalar and axial vector dipolar form factor (FFs) which has a form
\be 
g_{X_D}(k^2)=g_{X_D} \frac{(k^2-m^2_q)(1-x)^2}{(\bfk^2+L_{X_D}^2)^2} \,, 
\ee 
with $g_{X_D}$ as a coupling constant. On substituting the Dirac spinors and scalar vertex form from Eq. (\ref{scalar}) in Eq. (\ref{ScalarWfn}), the light-cone wave functions for a scalar diquark can be written as
\be 
\psi^{\Uparrow X}_\uparrow (x,\bfk) &=& (m_q+x M_X) \, \phi_X /x \, , \nonumber \\
\psi^{\Uparrow X}_\downarrow (x,\bfk) &=& -(k_1 +i k_2)  \phi_X /x \, , \nonumber \\
\psi^{\Downarrow X}_\uparrow (x,\bfk) &=& -\psi^{\Uparrow \ast}_\downarrow (x,\bfk) \, , \nonumber \\
\psi^{\Downarrow X}_\downarrow (x,\bfk) &=& \psi^{\Uparrow \ast}_\uparrow (x,\bfk) \, .
\ee 
Different choices are available for summing over all polarizations $d^{\mu \nu}=\sum_{\lambda_a} \epsilon^{\mu\ast}_{\lambda_a}  \epsilon^{\nu\ast}_{\lambda_a}$ in Refs. \cite{Bacchetta:2008af}. We have adopted $d^{\mu \nu}(P-k)=-g^{\mu \nu}$ in the present work. On considering transverse polarization with $\lambda_a=\pm$ and substituting the axial vector diquark vertex from Eq. (\ref{vector}) in Eq. (\ref{VectorWfn}), the light-cone wave functions for axial vector diquarks with transversely polarized diquark can be written as
\be 
\psi^{\Uparrow X}_{\uparrow +}(x,\bfk) &=& \frac{k_1 -i k_2}{1-x} \, \phi_X /x \, , \nonumber \\
\psi^{\Uparrow X}_{\uparrow -} (x,\bfk) &=& -x \frac{k_1 +i k_2}{1-x}  \phi_X /x \, , \nonumber \\
\psi^{\Uparrow X}_{\downarrow +} (x,\bfk) &=&(m_q+x M_X) \, \phi_X /x \, , \nonumber \\
\psi^{\Uparrow X}_{\downarrow -} (x,\bfk) &=& 0 \, , \nonumber \\
\psi^{\Downarrow X}_{\uparrow +}(x,\bfk) &=& 0 \, , \nonumber \\
\psi^{\Downarrow X}_{\uparrow -} (x,\bfk) &=& - \psi^{\Uparrow \ast}_{\downarrow +} (x,\bfk) \, , \nonumber \\
\psi^{\Downarrow X}_{\downarrow +} (x,\bfk) &=&\psi^{\Uparrow \ast}_{\uparrow -} (x,\bfk) \, , \nonumber \\
\psi^{\Downarrow X}_{\downarrow -} (x,\bfk) &=& \psi^{\Uparrow \ast}_{\uparrow +}(x,\bfk) \, .
\ee
For longitudinally polarized diquark with $\lambda_a=0$, the light-cone wave functions for axial vector diquarks  can be written as  
\be 
\psi^{\Uparrow X}_{\uparrow 0}(x,\bfk) &=& \frac{\bfk^2-x M_a^2-m_q M_X (1-x)^2}{\sqrt{2}(1-x)M_a} \, \phi_X /x \, , \nonumber \\
\psi^{\Uparrow X}_{\downarrow 0} (x,\bfk) &=& \frac{(m_q+M_X)(k_1 +i k_2)}{\sqrt{2}M_a}  \phi_X /x \, , \nonumber \\
\psi^{\Downarrow X}_{\uparrow 0} (x,\bfk) &=&\psi^{\Uparrow \ast}_{\downarrow 0} (x,\bfk) \, , \nonumber \\
\psi^{\Downarrow X}_{\downarrow 0} (x,\bfk) &=& - \psi^{\Uparrow \ast}_{\uparrow 0}(x,\bfk) \, .
\ee
For the completeness relation $d^{\mu \nu}(P-k)=-g^{\mu \nu}$, related to the summing over all polarization, the light-cone wave functions for an unphysical timelike polarized diquark can be written as
\be 
\psi^{\Uparrow X}_{\uparrow \mathcal{T}}(x,\bfk) &=& \frac{\bfk^2+x M_a^2-m_q M_X (1-x)^2}{\sqrt{2}(1-x)M_a} \, \phi_X /x \, , \nonumber \\
\psi^{\Uparrow X}_{\downarrow \mathcal{T}} (x,\bfk) &=& \frac{(m_q+M_X)(k_1 +i k_2)}{\sqrt{2}M_a}  \phi_X /x \, , \nonumber \\
\psi^{\Downarrow X}_{\uparrow \mathcal{T}} (x,\bfk) &=&\psi^{\Uparrow \ast}_{\downarrow 0} (x,\bfk) \, , \nonumber \\
\psi^{\Downarrow X}_{\downarrow \mathcal{T}} (x,\bfk) &=& - \psi^{\Uparrow \ast}_{\uparrow 0}(x,\bfk) \, .
\ee
The contribution of timelike polarization of diquark state adds up with an overall negative sign. The momentum space wave function $\phi_X$ written in above mentioned wave functions can be expressed as
\be 
\phi_X =- \frac{g_{X_D}}{\sqrt{1-x}} \frac{x (1-x)^2}{[\bfk+L^2_D]^2} \, ,
\ee 
with $L^2_{X_D}=x M^2_D+(1-x) \Lambda^2_X-x(1-x)M^2_X$. Here, $M_D$ is the mass of a diquark and $\Lambda_X$ is an appropriate cut-off to avoid the singularity and satisfy $M_D>M_X-\Lambda_X$ \cite{Jakob:1997wg}.

\section{Numerical Parameters \label{secnumpar}}
The input parameters used in our calculations are the mass of a baryon $M_{X}$, mass of a quark $m_{q}$ and mass of a spectator diquark $M_D$. Value of cut-off parameter $\Lambda_X$ has been chosen to fit the available EMFFs  data. Comparison with data \cite{CSSM:2014knt, Shanahan:2014cga, Cates:2011pz} has been demonstrated in next section and values of cut-off parametrs used in our calculation are $\Lambda_p=0.8$, $\Lambda_{\Sigma}=1.1$ and $\Lambda_{\Xi}=1.15$. Following Ref. \cite{Jakob:1997wg} for proton, particle data group and Ref. \cite{Zhang:2016qqg} for other particles, the values of the masses of baryons are summarized in Table \ref{tab_bmass} whereas the masses of  quarks and diquarks  are summarized in \ref{tab_qmass}.  \par 
\begin{table}[h]
	\centering
	\begin{tabular}{|c|c|c|c|c|c|}
		\hline
		$\text{Particle $(X)$}  $~~&~~$ p  $~~&~~$ \Sigma^{+} $~~&~~$ \Sigma^{-} $~~&~~$ \Xi^{o}  $~~&~~$  \Xi^{-} $\\
		\hline
		$\text{Mass, $M_{X}$ (GeV)} $~~&~~$ 0.938 $~~&~~$ 1.189 $~~&~~$ 1.197 $~~&~~$ 1.314 $~~&~~$ 1.312 $ \\
		\hline
	\end{tabular}
	\caption{Masses of members of octet baryons used in the present calculations.}
	\label{tab_bmass} 
\end{table}
\begin{table}[h]
	\centering
	\begin{tabular}{|c|c|c|c|c|c|c|c|}
		\hline
		$\text{Quantity}  $~~&~~$ m_{u/d}  $~~&~~$  m_s $~~&~~$ m_{\mathfrak{s}(uu/ud)} $~~&~~$ m_{\mathfrak{a}(uu/ud)} $~~&~~$ m_{\mathfrak{s}(us/ds)} $~~&~~$ m_{\mathfrak{a}(us/ds)} $~~&~~$ m_{ss} $ \\
		\hline
		$\text{Values (GeV)} $~~&~~$ 0.36 $~~&~~$ 0.48 $~~&~~$ 0.60 $~~&~~$ 0.80 $~~&~~$ 0.75 $~~&~~$ 0.95 $~~&~~$ 1.10 $ \\
		\hline
	\end{tabular}
	\caption{Masses of quarks and their combinations used in the present calculations.}
	\label{tab_qmass} 
\end{table}

\section{Electromagnetic Form Factors\label{SecEMFFs}}
EMFFs are the simplest baryonic structure observables. In a spin-$\frac{1}{2}$ system of a baryon, the EMFFs can be calculated from \cite{Brodsky:1980zm}
\be 
\langle P^\prime | J^\mu (0) |P \rangle = \bar{u} (P^\prime ) \big[ F_1(\Dp^2) \gamma^\mu  + F_2(\Dp^2) \frac{i}{2M_X} \sigma^{\mu \alpha} \Delta_\alpha \big] u(P) \, .
\ee 
Helicity conserving and helicity-flip vector current matrix elements of the $J^+$ current corresponds the Dirac and Pauli FFs respectively as
\be 
\bigg\langle P+\Delta ,\Uparrow \bigg| \frac{J^+(0)}{2 P^+}\bigg| P,\Uparrow \bigg\rangle &=& F_1(\Dp^2) \, , \nonumber \\
\bigg\langle P+\Delta ,\Uparrow \bigg| \frac{J^+(0)}{2 P^+}\bigg| P,\Downarrow \bigg\rangle &=& -(\Delta_1-i \Delta_2) \frac{F_2(\Dp^2)}{2M_X} \, .
\ee 
Dirac and Pauli FFs can be evaluated in the overlap form of light-cone wave functions from the electromagnetic current sandwiched between the initial and final baryon state as
\be 
F_1^{X_q}(\Dp^2) &=& \int \frac{dx~d^2 \bfk}{16 \pi^3} \sum_{\lambda_q, \lambda_D} \psi^{\Uparrow X \ast}_{\lambda_q}(x^\prime,\bfk^\prime) ~ \psi^{\Uparrow X}_{\lambda_q}(x,\bfk) \, , \nonumber \\ 
F_2^{X_q}(\Dp^2) &=& \int \frac{dx~d^2 \bfk}{16 \pi^3} \sum_{\lambda_q, \lambda_D} \psi^{\Uparrow X \ast}_{\lambda_q}(x^\prime,\bfk^\prime) ~ \psi^{\Downarrow X}_{\lambda_q}(x,\bfk) \, ,
\ee 
where $x^\prime$ and and $\bfk^\prime$ correspond to the final state longitudinal momentum fraction and transverse momentum carried by an active quark respectively and can be written as 
\begin{align}
x^\prime=x\, , && \bfk^\prime=\bfk+(1-x^\prime)\Dp \, .
\end{align}
\begin{figure*}
	\centering
	\begin{minipage}[c]{0.98\textwidth}
		(a)\includegraphics[width=7.5cm]{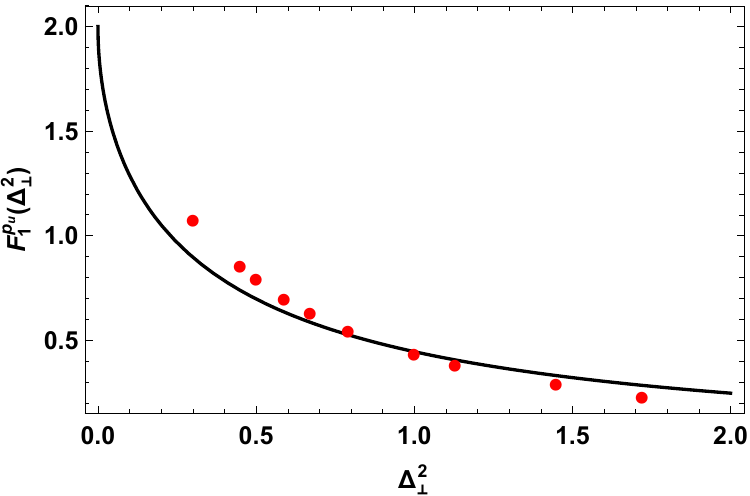}
		\hspace{0.03cm}
		(b)\includegraphics[width=7.5cm]{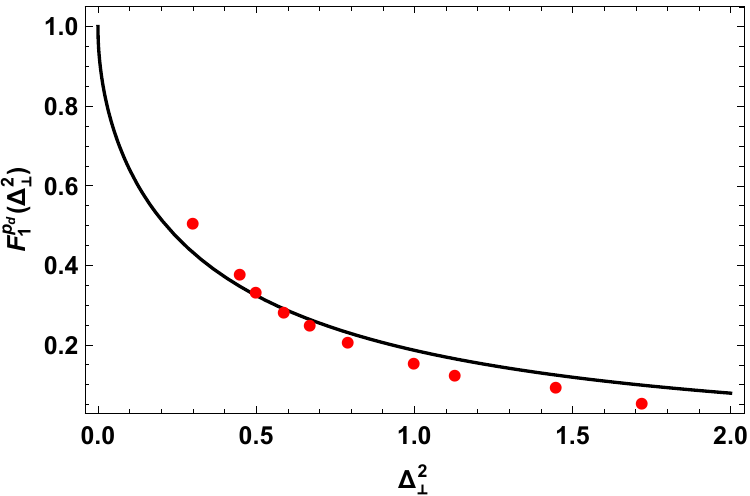}
		\hspace{0.03cm}	
		(c)\includegraphics[width=7.5cm]{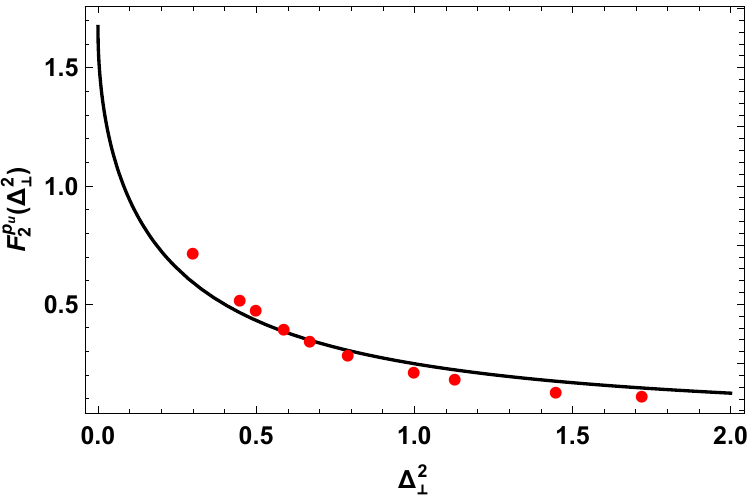}
		\hspace{0.03cm}
		(d)\includegraphics[width=7.5cm]{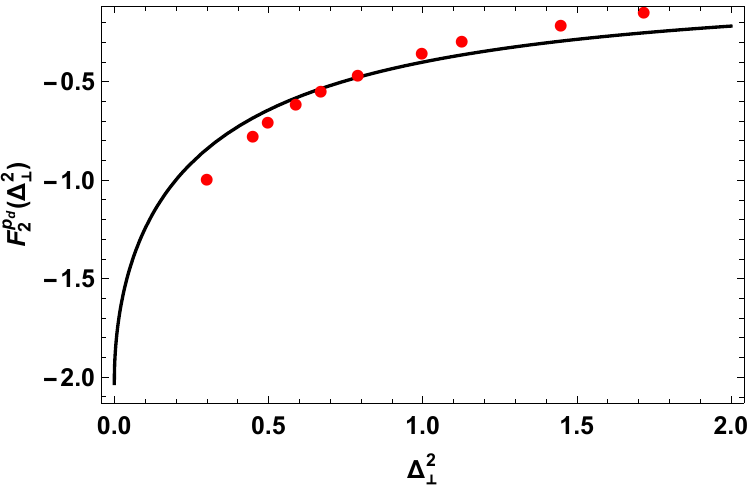}
		\hspace{0.03cm}	
	\end{minipage}
	\caption{\label{fig1FFs} (Color online) Dirac and Pauli form factors for constituent valence quarks of proton in first and second row respectively. The left side corresponds to the $u$ quarks and the right side to the $d$ quarks. Data points have been taken from Ref. \cite{Cates:2011pz}.}
\end{figure*} 
Following the constraints, $\sum_i x_{i}=1$ and $\sum_i k_{i}=0$, one can also define the longitudinal momentum fraction and transverse momentum of a spectator diquark. By considering the charge symmetry between isospin partners $u$ and $d$ quark flavors, one can write nucleon FFs in terms of its constituent quark flavors as 
\be 
F_j^{p(n)}=e_u F_j^{u(d)} + e_d F_j^{d(u)} \, ,
\label{FF}
\ee 
where $j=1,2$ correspond to Dirac and Pauli FFs respectively. In the similar manner, one can write the FFs of other members of octet baryon in terms of their constituent quark flavors between isospin partners. The Pauli and Dirac FFs for proton has to be normalized using $F_1^p(0)=1$ and $F_2^p(0)=\kappa^p=1.793$ constants. Whereas the normalization factors for its constituent quark flavors are defined as $F_1^{p_u}(0)=2$ and $F_1^{p_d}(0)=1$, because Dirac FFs for zero momentum transfer correspond to the number of quark flavors. Similarly, Pauli FFs present the anomalous magnetic moment at zero momentum transfer and it constraints the normalization condition for the constituent quark flavors as  $F_2^{p_u}(0)=\kappa^p_u=1.673$ and $F_2^{p_d}(0)=\kappa^p_d=-2.033$. Following a similar approach, one can generalize the normalization conditions for the constituent quarks as $F_1^{X_q}(0)=n_q^X$ and  $F_2^{X_q}(0)=\kappa^X_q$, where $n_q^X$ and $\kappa^X_q$ denote the number of $q$ flavored quarks in $X$ baryon and their anomalous magnetic moment respectively.  For the ease of interpretation, we have used the normalized functions for further calculations which can be interpreted as $f_{norm}^{X_q}=(N^2_{X_q}/g^2_{X_D}) f^{X_q}$, with $N_{X_q}$ as the normalization constant. The values $F_1^{X_q}(0)$ and $F_2^{X_q}(0)$ required to normalize the function $f^{X_q}$ for the other members of octet baryons can be computed in accordance with $\mu_X=G_M^X(Q^2=0)=F_1^X(0)+F_2^X(0)$ and Eq. \ref{FF}. The constants to normalize the functions at zero momentum transfer have been tabulated in Table \ref{tab_Norm}.
\begin{table}[h]
	\centering
	\begin{tabular}{|c|c|c|c|c|}
		\hline
		$\text{Quark flavor}  $~~&~~$ F_1^{\Sigma}(0) $~~&~~$ F_2^{\Sigma}(0) $~~&~~$ F_1^{\Xi}(0)  $ ~~&~~$ F_2^{\Xi}(0) ~~$ \\
		\hline
		$\text{$u/d$} $~~&~~$ 2 $~~&~~$ 1.618 $~~&~~$ 1 $~~&~~$ -1.599 $  \\
		\hline
		$\text{$s$} $~~&~~$ 1 $~~&~~$ -1.138 $~~&~~$ 2 $~~&~~$ 0.551 $ \\
		\hline
	\end{tabular}
	\caption{Values of Dirac and Pauli FFs at zero momentum transfer.}
	\label{tab_Norm} 
\end{table}

\begin{table}[h]
	\centering
\begin{tabular}{|c|c c|c c|c c|c c|}
	\hline
	$~ Q^2 ~$ & \multicolumn{2}{c|}{$F_1^{\Sigma^+_u}$}  & \multicolumn{2}{c|}{$F_1^{\Sigma^+_s}$} & \multicolumn{2}{c|}{$F_2^{\Sigma^+_u}$}  & \multicolumn{2}{c|}{$F_2^{\Sigma^+_s}$}  \\
	\cline{2-9}
    $~$ (GeV$^2$) $~$ & $~$ Lattice  $~$ & $~$ Our result  $~$ & $~$ Lattice  $~$ & $~$ Our result $~$ & $~$ Lattice  $~$ & $~$ Our result  $~$ & $~$ Lattice  $~$ & $~$ Our result $~$ \\
	\hline
	$~ 0.26 ~$ & ~$~ 1.367 ~$~ & ~$~ 1.218 ~$~ & ~$~ 0.655 ~$~ & ~$~ 0.566 ~$~ & ~$~ 0.915 ~$~ & ~$~ 0.826 ~$~ & ~$~ -1.076 ~$~ & ~$~ -0.626 ~$~ \\
	\hline
	$~ 0.50 ~$ & ~$~ 1.053 ~$~ & ~$~ 0.981 ~$~ & ~$~ 0.473 ~$~ & ~$~ 0.434 ~$~ & ~$~ 0.714 ~$~ & ~$~ 0.623 ~$~ & ~$~ -0.815 ~$~ & ~$~ -0.487 ~$~ \\
	\hline
	$~ 0.73 ~$ & ~$~ 0.864 ~$~ & ~$~ 0.832 ~$~ & ~$~ 0.359 ~$~ & ~$~ 0.352 ~$~ & ~$~ 0.555 ~$~ & ~$~ 0.504 ~$~ & ~$~ -0.633 ~$~ & ~$~ -0.404 ~$~ \\
	\hline
	$~ 0.94 ~$ & ~$~ 0.734 ~$~ & ~$~ 0.732 ~$~ & ~$~ 0.297 ~$~ & ~$~ 0.297 ~$~ & ~$~ 0.414 ~$~ & ~$~ 0.429 ~$~ & ~$~ -0.529 ~$~ & ~$~ -0.350 ~$~ \\
	\hline
	$~ 1.14 ~$ & ~$~ 0.624 ~$~ & ~$~ 0.657 ~$~ & ~$~ 0.238 ~$~ & ~$~ 0.257 ~$~ & ~$~ 0.343 ~$~ & ~$~ 0.374 ~$~ & ~$~ -0.442 ~$~ & ~$~ -0.310 ~$~ \\
	\hline
	$~ 1.33 ~$ & ~$~ 0.554 ~$~ & ~$~ 0.598 ~$~ & ~$~ 0.198 ~$~ & ~$~ 0.225 ~$~ & ~$~ 0.296 ~$~ & ~$~ 0.333 ~$~ & ~$~ -0.368 ~$~ & ~$~ -0.279 ~$~ \\
	\hline
\end{tabular}
\caption{Juxtapose for the results of Dirac and Pauli form factors for $u$ and $s$ constituent quark flavors for $\Sigma$ baryon with lattice results \cite{CSSM:2014knt}.}
\label{TabSigmaFFs} 
\end{table} 
\begin{table}[h]
	\centering
	\begin{tabular}{|c|c c|c c|c c|c c|}
		\hline
		$~ Q^2 ~$ & \multicolumn{2}{c|}{$F_1^{\Xi^o_u}$}  & \multicolumn{2}{c|}{$F_1^{\Xi^o_s}$} & \multicolumn{2}{c|}{$F_2^{\Xi^o_u}$}  & \multicolumn{2}{c|}{$F_2^{\Xi^o_s}$} \\
		\cline{2-9}
		$~$ (GeV$^2$) $~$ & $~$ Lattice  $~$ & $~$ Our result  $~$ & $~$ Lattice  $~$ & $~$ Our result $~$ & $~$ Lattice  $~$ & $~$ Our result  $~$ & $~$ Lattice  $~$ & $~$ Our result $~$ \\
		\hline
		$~ 0.26 ~$ & ~$~ 1.409 ~$~ & ~$~ 1.199 ~$~ & ~$~ 0.628 ~$~ & ~$~ 0.610 ~$~ & ~$~ -1.082 ~$~ & ~$~ -0.924 ~$~ & ~$~ 0.892 ~$~ & ~$~ 0.285 ~$~ \\
		\hline
		$~ 0.50 ~$ & ~$~ 1.103 ~$~ & ~$~ 0.961 ~$~ & ~$~ 0.441 ~$~ & ~$~ 0.483 ~$~ & ~$~ -0.795 ~$~ & ~$~ -0.734 ~$~ & ~$~ 0.684 ~$~ & ~$~ 0.216 ~$~ \\
		\hline
		$~ 0.73 ~$ & ~$~ 0.911 ~$~ & ~$~ 0.813 ~$~ & ~$~ 0.331 ~$~ & ~$~ 0.401 ~$~ & ~$~ -0.623 ~$~ & ~$~ -0.618 ~$~ & ~$~ 0.546 ~$~ & ~$~ 0.176 ~$~ \\
		\hline
		$~ 0.95 ~$ & ~$~ 0.792 ~$~ & ~$~ 0.709 ~$~ & ~$~ 0.272 ~$~ & ~$~ 0.344 ~$~ & ~$~ -0.501 ~$~ & ~$~ -0.537 ~$~ & ~$~ 0.430 ~$~ & ~$~ 0.149 ~$~ \\
		\hline
		$~ 1.14 ~$ & ~$~ 0.677 ~$~ & ~$~ 0.635 ~$~ & ~$~ 0.217 ~$~ & ~$~ 0.302 ~$~ & ~$~ -0.424 ~$~ & ~$~ -0.480 ~$~ & ~$~ 0.352 ~$~ & ~$~ 0.130 ~$~ \\
		\hline
		$~ 1.33 ~$ & ~$~ 0.594 ~$~ & ~$~ 0.578 ~$~ & ~$~ 0.179 ~$~ & ~$~ 0.269 ~$~ & ~$~ -0.354 ~$~ & ~$~ -0.436 ~$~ & ~$~ 0.306 ~$~ & ~$~ 0.116 ~$~ \\
		\hline
	\end{tabular}
	\caption{Juxtapose for the results of Dirac and Pauli form factors for $u$ and $s$ constituent quark flavors for $\Xi$ baryon with lattice results \cite{CSSM:2014knt}.}
	\label{TabXiFFs} 
\end{table} 
The cut-off parameter $\Lambda_X$ for the constituent quark flavors of proton, defined in the expression of $L_{X_D}$, has been fixed by relating the outcomes of FFs with the available data. For $\Sigma$ and $\Xi$, the cut-off parameter $\Lambda_X$ has been fixed using the available lattice data \cite{CSSM:2014knt}. Plots corresponding to the Dirac and Pauli FFs for $u$ as well as $d$ quark flavors for the case of proton have been portrayed in Fig. \ref{fig1FFs}. The Dirac and Pauli FFs for $\Sigma$ and $\Xi$ follow a similar behavior and also since there is no data available for them, we have presented in Fig. \ref{fig1FFs} only the case of proton. However, the computed values of $\Sigma^+$ and $\Xi^0$ FFs at different values of $Q^2 (=-\Delta^2=\Dp^2)$ have been presented in Tables \ref{TabSigmaFFs} and \ref{TabXiFFs} respectively. The available lattice data \cite{CSSM:2014knt} of the FFs have also been presented in the tables. It is important to mention here that due to similar characteristics of FFs among isospin partners of $\Sigma$ ($\Sigma^+$ and $\Sigma^-$) and $\Xi$ ($\Xi^0$ and $\Xi^-$), we have presented the generalized distributions for one of them in case of FFs as well as for other distribution functions discussed in forthcoming sections. From the tables, it is clear that Dirac FFs for $u$ as well as $s$ quark flavors as well as the Pauli FF of $u$ quark flavor is in good agreement with lattice data. However, Pauli FF corresponding to $s$ quark flavor comes out to be different. This difference may be attributed to the baseline of the model assumptions. In our calculations, the Pauli FFs satisfy $\mu_X=G_M^X(Q^2=0)=F_1^X(0)+F_2^X(0)$ and Eq. (\ref{FF}) by considering charge symmetry between $u$ and $d$ flavor quark among isospin partners.

\section{Generalized Parton Distribution Functions\label{SecGPDs}}
\begin{figure*}
	\centering
	\begin{minipage}[c]{0.98\textwidth}
		\includegraphics[width=7.5cm]{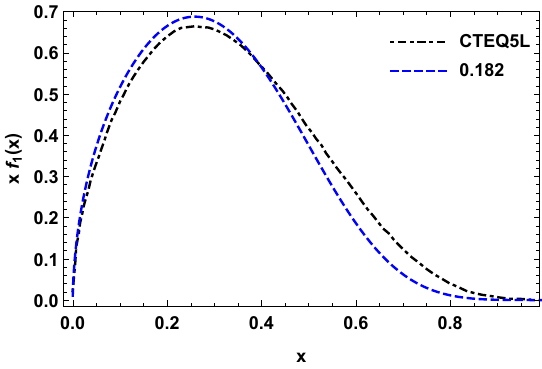}
		\hspace{0.03cm}			
	\end{minipage}
	\caption{\label{fig2PDFComp} (Color online) Model calculation of $x f_{1}(x)$ (dashed line) compared to the CTEQ$5$L parametrization \cite{Lai:1999wy} (dot dashed line) at 1 GeV$^{2}$.}
\end{figure*}

In the light-cone gauge $A^+=0$, the off-forward matrix elements of the light-cone bilinear vector current provides chiral-even unpolarized hadron GPDs as \cite{Dahiya:2007is, Meissner:2009ww}
\be
F^{\gamma^+}_{\lambda_X^\prime \lambda_X}=\frac{1}{2}\int \frac{dy^{-}}{2 \pi} e^{ix P^{+}y^{-}} \bigg\langle P^{\prime},\lambda_X^\prime\bigg|\bar{\psi}\bigg(\frac{-y}{2}\bigg) \,\gamma^{+} \,\psi\bigg(\frac{y}{2}\bigg)\bigg|P,\lambda_X\bigg\rangle \bigg|_{y^{+}=0, \bf{y_{\perp}}=0} \, , 
\ee
which is parameterized using
\be
F^{\gamma^+}_{\lambda_X^\prime \lambda_X}= \frac{1}{2 \bar{P^+}}\bar{u}(P^\prime)\bigg[H_{X}^{q}(x,0,\Dp) \,\gamma^{+}+E_{X}^{q}(x,0,\Dp) \,\frac{i \sigma^{+\alpha}(-\Delta_{\alpha})}{2M_X}\bigg]u(P) \, ,
\ee 
where $\alpha$ corresponds to the transverse indices. By making the use of spin-$1/2$ Dirac spinors, quark field operators and baryon states, one can write the GPDs in terms of light-cone wave functions as \cite{Hwang:2007tb}
\be 
\frac{\sqrt{1-\zeta}}{1-\frac{\zeta}{2}}H^{X_q}(x,\zeta,\Dp) - \frac{\zeta^2}{4 (1-\frac{\zeta}{2}) \sqrt{1-\zeta}} E^{X_q}(x,\zeta,\Dp) &=& \int \frac{dx~d^2 \bfk}{16 \pi^3} \sum_{\lambda_q, \lambda_D} \psi^{\Uparrow X \ast}_{\lambda_q}(x^\prime,\bfk^\prime) ~ \psi^{\Uparrow X}_{\lambda_q}(x,\bfk) \, , \nonumber \\ 
\frac{1}{\sqrt{1-\zeta}}\frac{-\Delta_1+i \Delta_2}{2M_X}E^{X_q}(x,\zeta,\Dp) &=& \int \frac{dx~d^2 \bfk}{16 \pi^3} \sum_{\lambda_q, \lambda_D} \psi^{\Uparrow X \ast}_{\lambda_q}(x^\prime,\bfk^\prime) ~ \psi^{\Downarrow X}_{\lambda_q}(x,\bfk) \, , \nonumber \\  
&& \ee 
where $x^\prime$ and and $\bfk^\prime$ correspond to the final state longitudinal momentum fraction and transverse momentum carried by an active quark respectively and can be expressed  as 
\be
	x^\prime&=&\frac{x-\zeta}{1-\zeta}\, , \nonumber \\ 
	\bfk^\prime&=&\bfk+(1-x^\prime)\Dp \, .
\ee
The explicit expressions corresponding to scalar and axial vector diquarks for spin non-flip GPDs are as follows
\be 
\frac{\sqrt{1-\zeta}}{1-\frac{\zeta}{2}} H^{X_q}_{\mathfrak{s}}(x,\zeta,\Dp)&=&\int \frac{d^2\bfk}{16 \pi^3} ~ \bigg[ \biggl\{ (m_q+x M_X)~(m_q+x^\prime M_X)+\bfk^2+\frac{1-x}{1-\zeta} ~ \bfk \cdot \Dp \biggr\} ~ \frac{\phi^\prime_\mathfrak{s} \phi_\mathfrak{s}}{x^\prime x} 
\nonumber \\ &+& \frac{\zeta^2}{4 (1-\frac{\zeta}{2}) \sqrt{1-\zeta}} E^{X_q}_\mathfrak{s}(x,\zeta,\Dp^2) \bigg] \, , \label{GPD_H} \\
\frac{\sqrt{1-\zeta}}{1-\frac{\zeta}{2}} H^{X_q}_\mathfrak{a} (x,\zeta,\Dp)&=& \int \frac{d^2\bfk}{16 \pi^3} ~ \bigg[ \biggl\{ (1-x^\prime)(1-x)(\bfk^2+m_q^2+(x+x^\prime) ~ m_q M_X +x^\prime x M^2_X) 
\nonumber \\ &+&  \big(x^\prime(1-x^2)+x(1-x^\prime)^2\big) ~ m_q M_X - (1+x^\prime x) (1-x^\prime) ~ \bfk \cdot \Dp 
\nonumber \\ &-& x ~ \big((1-x^\prime)^2 \Dp^2 - 2(1-x^\prime) ~ \bfk \cdot \Dp \big)
\biggr\} ~ \frac{\phi^\prime_\mathfrak{a} \phi_\mathfrak{a}}{x x^\prime (1-x) (1-x^\prime)} \nonumber \\ &+& \frac{\zeta^2}{4 (1-\frac{\zeta}{2}) \sqrt{1-\zeta}} E^{X_q}_\mathfrak{a}(x,\zeta,\Dp^2) \bigg] \, .
\ee 
The unintegrated expressions for zero skewness and zero momentum transfer correspond to the expressions of unpolarized TMD, expressed in Appendix of Ref. \cite{Bacchetta:2008af}. 
Whereas the explicit expressions corresponding to scalar and axial vector diquarks for spin flip GPDs are as follows
\be 
\frac{1}{\sqrt{1-\zeta}} E^{X_q}_\mathfrak{s}(x,\zeta,\Dp) &=& \int \frac{d^2\bfk}{16 \pi^3} ~ 2M_X ~ \bigg[ (m_q+x M_X)(1-x)-\frac{\zeta(-1+x)}{1-\zeta} M_X \frac{\bfk}{\Dp} \bigg] ~ \frac{\phi^\prime_\mathfrak{s} \phi_\mathfrak{s}}{x x^\prime} \, ,\\
\frac{1}{\sqrt{1-\zeta}} E^{X_q}_\mathfrak{a}(x,\zeta,\Dp) &=& \int \frac{d^2\bfk}{16 \pi^3} ~ 2M_X ~ \bigg[-x(m_q+x^\prime M_X) + \frac{\zeta}{1-x} (m_q+xx^\prime M_X) \frac{\bfk}{\Dp} 
\nonumber \\ &-& (m_q+M_X) \biggl\{x+\frac{\zeta}{1-x} \frac{\bfk}{\Dp} \biggr\} \bigg] ~ \frac{\phi^\prime_\mathfrak{a} \phi_\mathfrak{a}}{x x^\prime} \, . 
\ee
In the forward limit with zero skewness, GPDs are reduced to PDFs which contain information about finding a particular quark flavor with longitudinal momentum fraction $x$. One can estimate the model scale by comparing the total momentum carried by the valence quarks with some parameterization results. In order to do so, we have evolved our result upto 1 GeV$^2$ to compare with CTEQ$5$L parameterization. Our model scale is at 0.182 GeV$^2$  for given value of $\Lambda=0.8$ which is different from the scale in Ref. \cite{Jakob:1997wg}. This is because we have chosen a cut-off value in order to get comparable FFs with the available data. The evolution is carried out using DGLAP equation and the comparison of our evolved PDF and CTEQ$5$L parameterization \cite{Lai:1999wy} is presented in Fig. \ref*{fig2PDFComp}. \par
\begin{figure*}
	\centering
	\begin{minipage}[c]{0.98\textwidth}
		(a)\includegraphics[width=7.5cm]{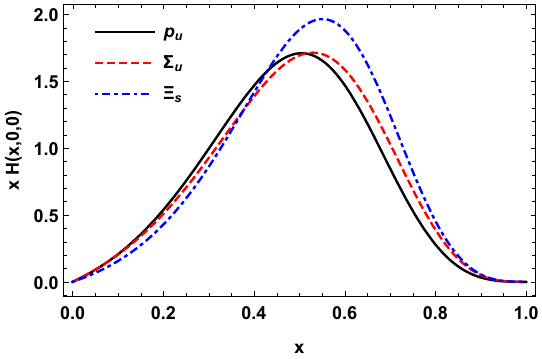}
		\hspace{0.03cm}
		(b)\includegraphics[width=7.5cm]{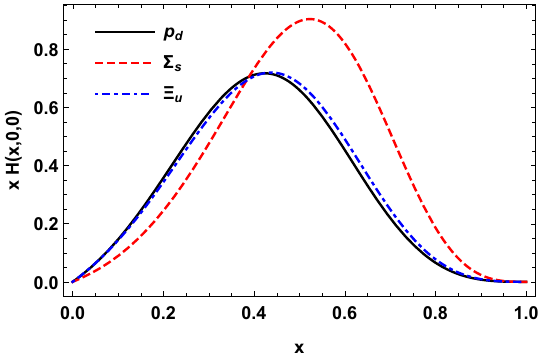}
		\hspace{0.03cm}			
	\end{minipage}
	\caption{ (Color online) Spin non-flip generalized parton distributions $x H(x,0,0)$ of valence quarks in proton, $\Sigma$ and $\Xi$ with zero skewness and zero transverse momentum transfer, usually termed as parton distribution functions, $f_{1}(x)=H(x,0,0)$.} \label{fig3}
\end{figure*} 

\begin{figure*}
	\centering
	\begin{minipage}[c]{0.98\textwidth}
		(a)\includegraphics[width=7.5cm]{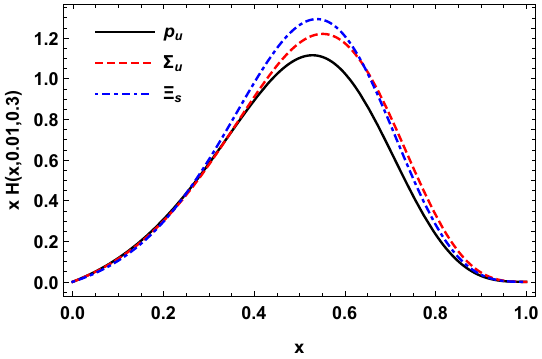}
		\hspace{0.03cm}
		(b)\includegraphics[width=7.5cm]{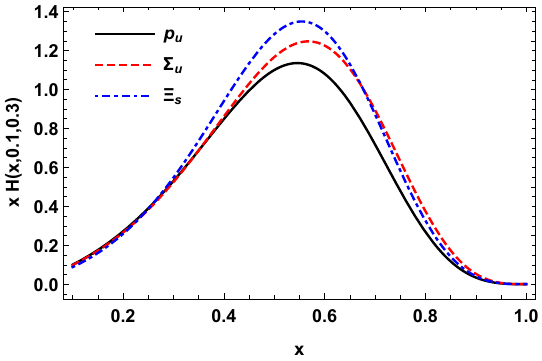}
		\hspace{0.03cm}
		(c)\includegraphics[width=7.5cm]{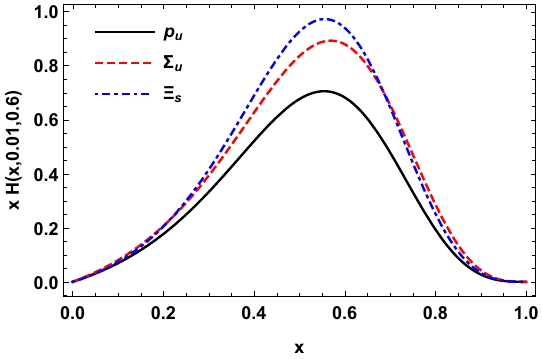}
		\hspace{0.03cm}
		(d)\includegraphics[width=7.5cm]{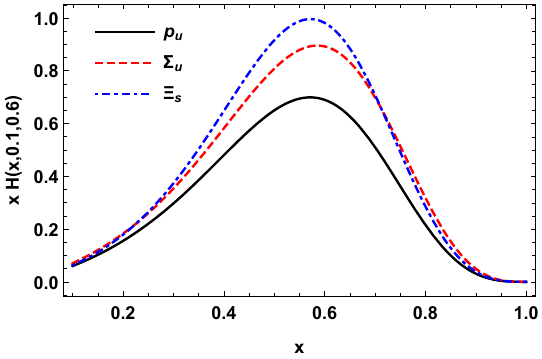}
		\hspace{0.03cm}				
	\end{minipage}
	\caption{\label{fig4GPDsHuus} (Color online) Spin non-flip generalized parton distributions for valence quark flavors in proton, $\Sigma$ and $\Xi$, having the contribution of both scalar and axial vector diquarks. In the first row, distributions correspond to the fixed transverse momentum transfer $\Dp=0.3$ for two different values of skewness parameter ($\zeta=0.01$ in Fig. (a) and $\zeta=0.1$ in Fig. (b)). In the second row, distributions correspond to the fixed transverse momentum transfer $\Dp=0.6$ for two different values of skewness parameter ($\zeta=0.01$ in Fig. (c) and $\zeta=0.1$ in Fig. (d)).}
\end{figure*} 
In order to portray the ability to find each quark flavor of $p$, $\Sigma$, and $\Xi$ with longitudinal momentum fraction $x$, their PDFs have been presented in Fig. \ref*{fig3}. Those valence quark distributions that account for the contribution of both scalar and axial vector diquarks are presented in Fig. \ref{fig3} (a). It is clear from the figure that $x=0.551$ is the most probable longitudinal momentum fraction carried by $s$ quark flavor of $\Xi$. Whereas $x=0.530$ and $x=0.505$ are the most probable values of longitudinal momentum fraction carried by $u$ quark flavor of $\Sigma$ and $p$ respectively. Further, the $s$ quark flavor of $\Xi$ carries more fraction of longitudinal momentum as compared to the $u$ quark  flavor in $\Sigma$ despite having the same diquark $us$ in both baryons. This suggests that heavier the quark, more fraction of longitudinal momentum is carried by it.  However, no prominent difference exists in the amplitudes and peak values of the longitudinal momentum fraction of $p$ and $\Sigma$. This may be attributed to the compensation of increased $us$ diquark mass by a larger parent $\Sigma$ baryon mass. Fig. \ref{fig3} (b) illustrates the valence quark distributions that include the contribution of solely axial vector diquarks. It is clear that the $s$ quark flavor in $\Sigma$, owing to its higher mass, carries more fraction of longitudinal momentum than the $u$ quark flavor in $p$, even though both baryons contain the same diquark $uu$. On the other hand, the $u$ quark flavor in $\Xi$ carries a smaller longitudinal momentum fraction $x$ than $s$ in $\Sigma$ due to the presence of heavier $ss$ diquark in $\Xi$ that carries a larger share of longitudinal momentum fraction as compared to a lighter $uu$ diquark in $\Sigma$. 

To analyze the dependence of skewness and transverse momentum transfer on spin non-flip GPDs of valence quark flavors that include both scalar and axial vector contributions, we have presented the distributions in Fig. \ref{fig4GPDsHuus} by fixing the transverse momentum transfer and varying the skewness parameter as well by fixing the  skewness parameter and varying the transverse momentum transfer. The general pattern of spin non-flip GPDs about carrying the longitudinal momentum fraction $x$ among constituent quark flavors of their parent baryons is similar to that in the case of zero skewness GPDs. The impact of change in skewness from $\zeta=0.01$ and $0.1$ can be analyzed by moving from the left to right panel of Fig. \ref{fig4GPDsHuus} for two distinct values of momentum transfer  $\Dp=0.3$ and $0.6$ in the first and second row, respectively. The peaks are found to be shifted to higher values of $x$ along with the increment in amplitudes with increasing skewness of the baryon. A similar kind of behavior was also predicted in Ref. \cite{Mondal:2015uha} for the case of proton. On the other hand, the effect of momentum transfer for  the value of $\Dp$ from $0.3$ to $0.6$ can be studied on moving from up to down for two fixed values of skewness $\zeta=0.01$ and $0.1$ in the left and right panels, respectively. The amplitude of these distributions follow the generic trend of falling down with an increase in the values of $\Dp$. However, the amplitude of distributions corresponding to $u$ quark flavor of $p$ is found to scale down more rapidly  with increasing $\Dp$ as compared to the strange baryons. 

\begin{figure*}
	\centering
	\begin{minipage}[c]{0.98\textwidth}
		(a)\includegraphics[width=7.5cm]{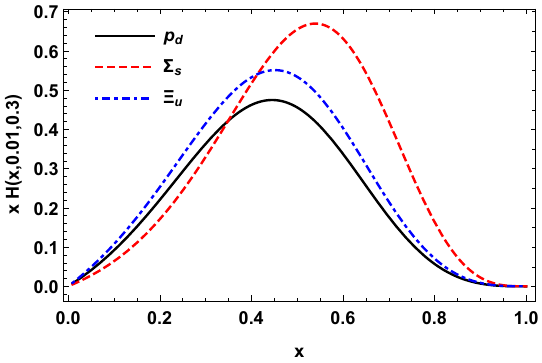}
		\hspace{0.03cm}
		(b)\includegraphics[width=7.5cm]{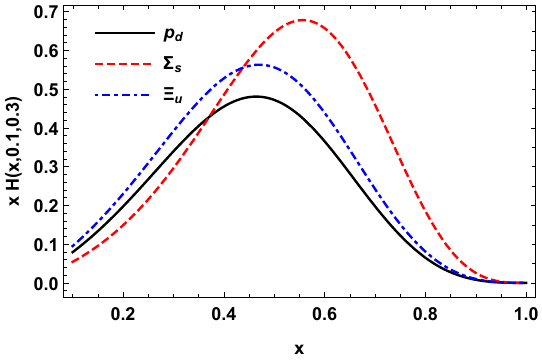}
		\hspace{0.03cm}
		(c)\includegraphics[width=7.5cm]{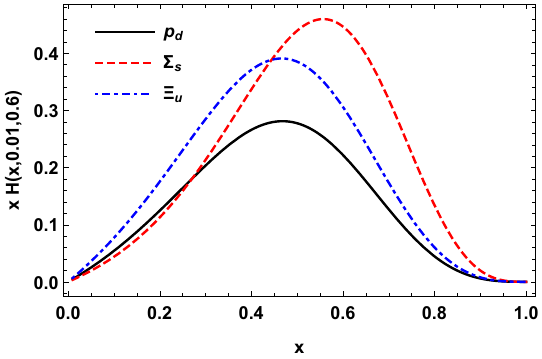}
		\hspace{0.03cm}
		(d)\includegraphics[width=7.5cm]{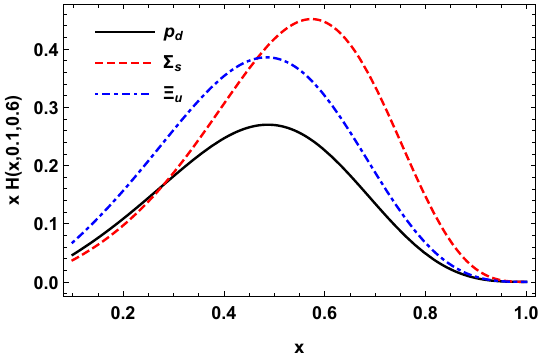}
		\hspace{0.03cm}				
	\end{minipage}
	\caption{\label{fig5GPDsHssu} (Color online) Spin non-flip generalized parton distributions for valence quark flavors, having the contribution of axial vector diquark only. In the first row, distributions correspond to the fixed transverse momentum transfer $\Dp=0.3$ for two different values of skewness parameter $\zeta=0.01$ in Fig. (a) and $\zeta=0.1$ in Fig. (b). In the second row, distributions correspond to the fixed transverse momentum transfer $\Dp=0.6$ for two different values of skewness parameter $\zeta=0.01$ in Fig. (c) and $\zeta=0.1$ in Fig. (d).}
\end{figure*} 
\begin{figure*}
	\centering
	\begin{minipage}[c]{0.98\textwidth}
		(a)\includegraphics[width=7.5cm]{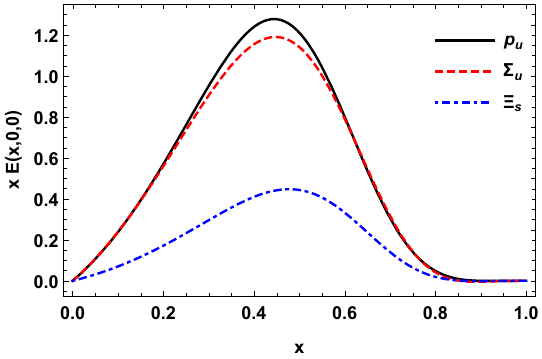}
		\hspace{0.03cm}
		(b)\includegraphics[width=7.5cm]{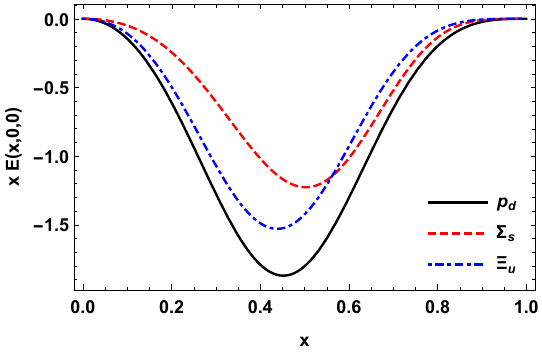}
		\hspace{0.03cm}			
	\end{minipage}
	\caption{\label{fig6} (Color online) Generalized parton distributions $x E(x,0,0)$ of valence quarks in proton, $\Sigma$ and $\Xi$ with zero skewness and zero transverse momentum transfer.}
\end{figure*}  
Fig. \ref{fig5GPDsHssu} represents the spin non-flip GPDs of valence quark flavors that include only the axial vector contribution to study their dependence of skewness and transverse momentum transfer. The over all behavior of skewed GPDs about carrying the longitudinal momentum fraction $x$ among constituent quark flavors of their parent baryons is similar to the non-skewed GPDs, presented in Fig. \ref{fig3} (b). The dependence of skewness corresponding to $\zeta=0.01$ and $0.1$ can be analyzed by moving from the left to right panels of Fig. \ref{fig5GPDsHssu} for two distinct values of momentum transfer  $\Dp=0.3$ and $0.6$ in the first and second row, respectively. The peaks are observed to shift towards higher $x$ with a slight increase in amplitudes as compared to the distribution which include contributions of both scalar and axial vector diquarks. The dependence of momentum transfer for $\Dp=0.3$ and $0.6$ can be examined on moving from up to down for two fixed values of skewness $\zeta=0.01$ and $0.1$ in the left and right panels, respectively. A significant amount of amplitude can be found for the case of strange baryons $\Sigma$ and $\Xi$ as compare to $p$ at larger values of $\Dp$. Otherwise, the general trend of decrease in the amplitude of distributions with increasing $\Dp$ is same.

In the forward limit, spin-flip GPD that accounts for the contribution of both scalar and axial vector diquarks are presented in Fig. \ref{fig6} (a). The trend of carrying a longitudinal momentum fraction is the same as that of spin non-flip GPD. We have found that the most probable longitudinal momentum fraction carried by $u$ quark flavor in $p$ and $\Sigma$ is $x=0.444$ and $0.447$ respectively. Whereas, for the $s$ quark flavor in $\Xi$, it is $x=0.477$. As the spin-flip GPD is related to anomalous magnetic moment, the amplitudes of different quark flavors of baryons depend on it and are found to be the least for the $s$ quark flavor in $\Xi$. The distributions corresponding to the contribution of axial vector diquark only for spin-flip GPDs are presented in \ref{fig6} (b). Negative distributions reflect negative value of anomalous magnetic moment of the constituent quark flavors. $\Xi$ baryon with the bulkiest diquark $ss$ provides the smallest share of longitudinal momentum fraction to it, which is $x=0.438$. Among $p$ and $\Sigma$, with lighter diquark $uu$, the $s$ quark flavor in $\Sigma$ carries more longitudinal momentum fraction $x=0.456$ as compared to  $u$ with $x=0.452$ in $p$. 
 
\begin{figure*}
	\centering
	\begin{minipage}[c]{0.98\textwidth}
		(a)\includegraphics[width=7.5cm]{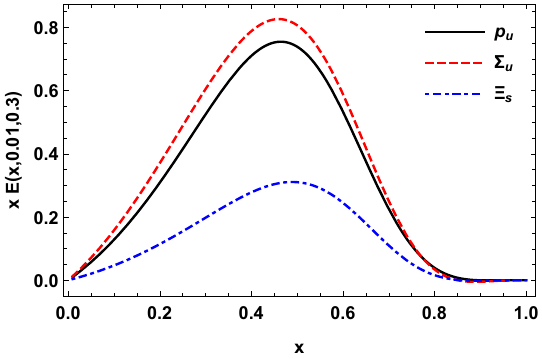}
		\hspace{0.03cm}
		(b)\includegraphics[width=7.5cm]{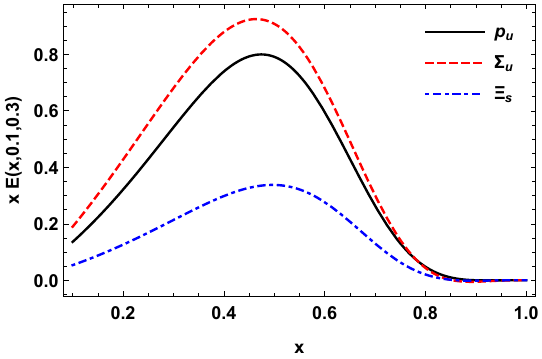}
		\hspace{0.03cm}	
		(c)\includegraphics[width=7.5cm]{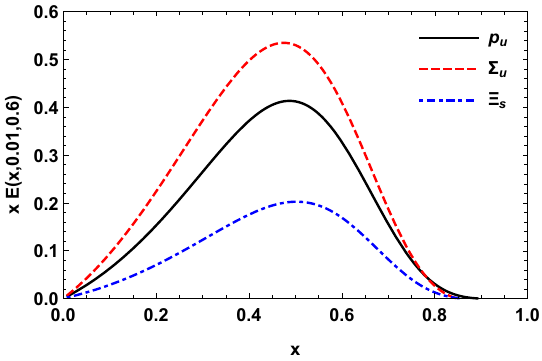}
		\hspace{0.03cm}
		(d)\includegraphics[width=7.5cm]{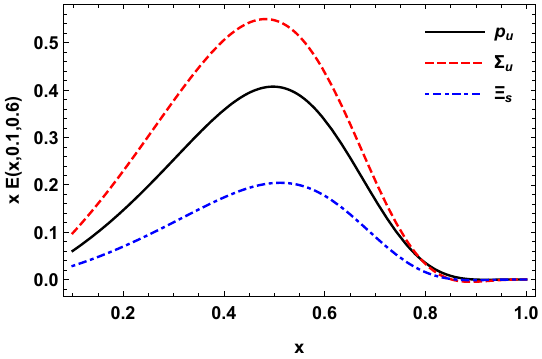}
		\hspace{0.03cm}		
	\end{minipage}
	\caption{\label{fig7Euus} (Color online) Spin flip generalized parton distributions for valence quark flavors, having the contribution of both scalar and axial vector diquarks. In the first row, distributions correspond to the fixed transverse momentum transfer $\Dp=0.3$ for two different values of skewness parameter $\zeta=0.01$ in Fig. (a) and $\zeta=0.1$ in Fig. (b). In the second row, distributions correspond to the fixed transverse momentum transfer $\Dp=0.6$ for two different values of skewness parameter $\zeta=0.01$ in Fig. (c) and $\zeta=0.1$ in Fig. (d).}
\end{figure*} 
To examine the influence of skewness and transverse momentum transfer on the spin-flip GPDs of valence quarks, including the contributions of both scalar and vector diquarks, we have presented the distributions in Fig. \ref{fig7Euus}. The influence of increased skewness from $\zeta=0.01$ to $0.1$ can be analyzed by moving from left to right panel for two different values of $\Dp=0.3$ and $0.6$ in the left and right panel respectively. The enhancement in peak amplitude and the shifting of peak value towards higher $x$ is observed. Whereas, the amplitude of the peaks scale down with the shifting of peak towards higher $x$ as the transverse momentum transfer is increased from $\Dp=0.3$ to $0.6$ on viewing from top to down for two fixed values of  skewness $\zeta=0.01$ and $0.1$ in the left and right panels respectively.  
\begin{figure*}
	\centering
	\begin{minipage}[c]{0.98\textwidth}
		(a)\includegraphics[width=7.5cm]{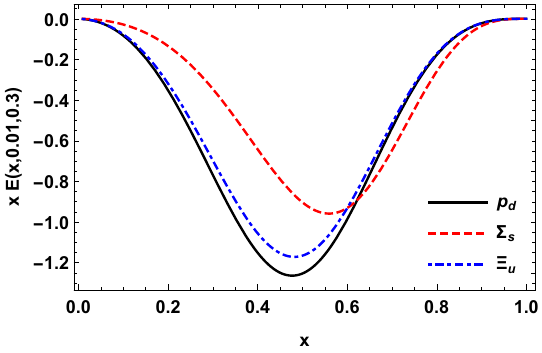}
		\hspace{0.03cm}
		(b)\includegraphics[width=7.5cm]{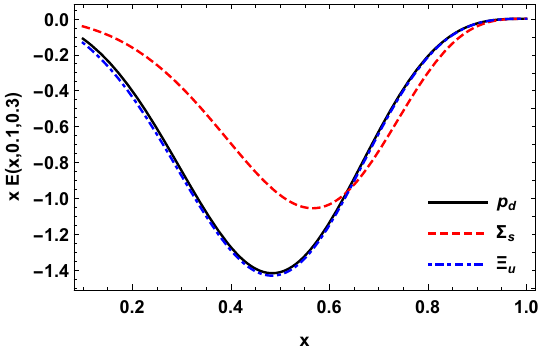}
		\hspace{0.03cm}	
		(c)\includegraphics[width=7.5cm]{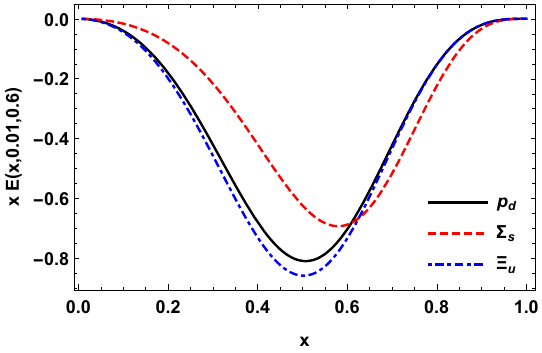}
		\hspace{0.03cm}
		(d)\includegraphics[width=7.5cm]{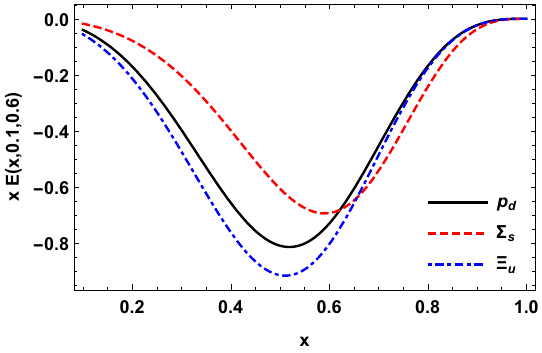}
		\hspace{0.03cm}		
	\end{minipage}
	\caption{\label{fig8Essu} (Color online) Spin-flip generalized parton distributions for valence quark flavors, having the contribution of axial vector diquark only. In the first row, distributions correspond to the fixed transverse momentum transfer $\Dp=0.3$ for two different values of skewness parameter $\zeta=0.01$ in Fig. (a) and $\zeta=0.1$ in Fig. (b). In the second row, distributions correspond to the fixed transverse momentum transfer $\Dp=0.6$ for two different values of skewness parameter $\zeta=0.01$ in Fig. (c) and $\zeta=0.1$ in Fig. (d).}
\end{figure*} 
\begin{figure*}
	\centering
	\begin{minipage}[c]{0.98\textwidth}
		(a)\includegraphics[width=7.5cm]{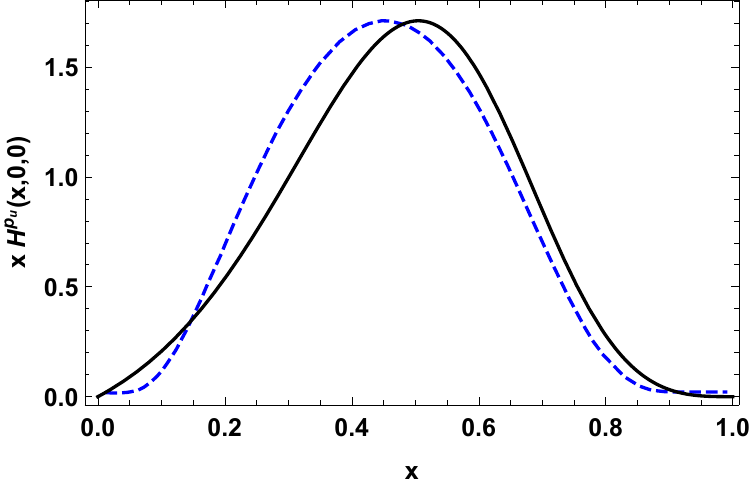}
		\hspace{0.03cm}
		(b)\includegraphics[width=7.5cm]{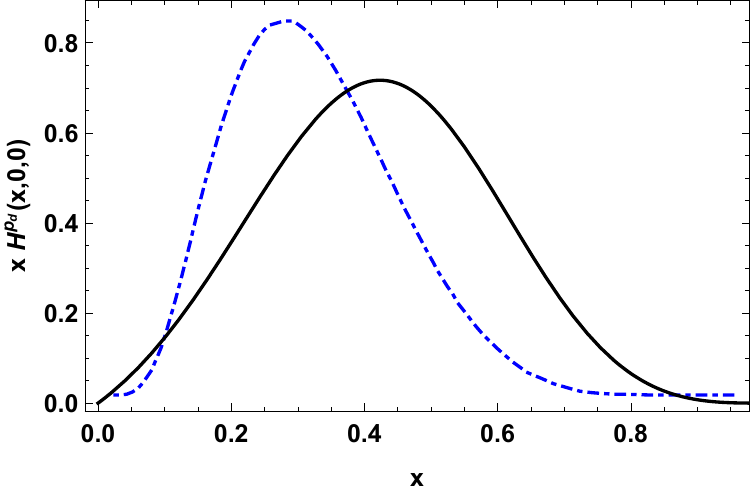}
		\hspace{0.03cm}	
		(c)\includegraphics[width=7.5cm]{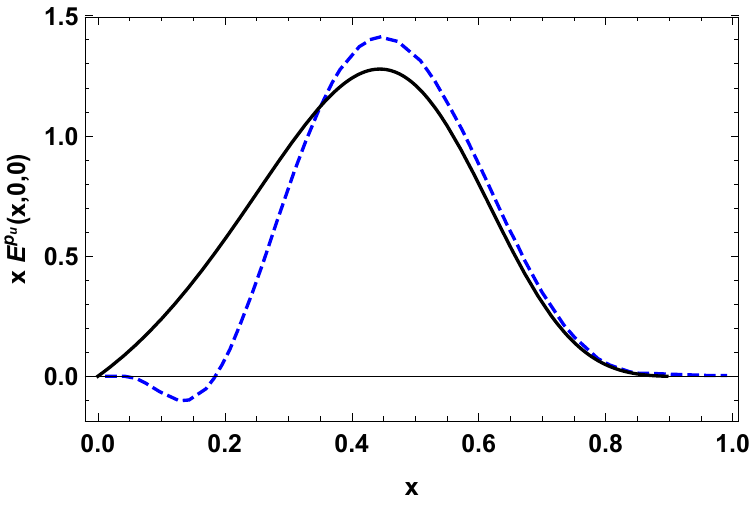}
		\hspace{0.03cm}
		(d)\includegraphics[width=7.5cm]{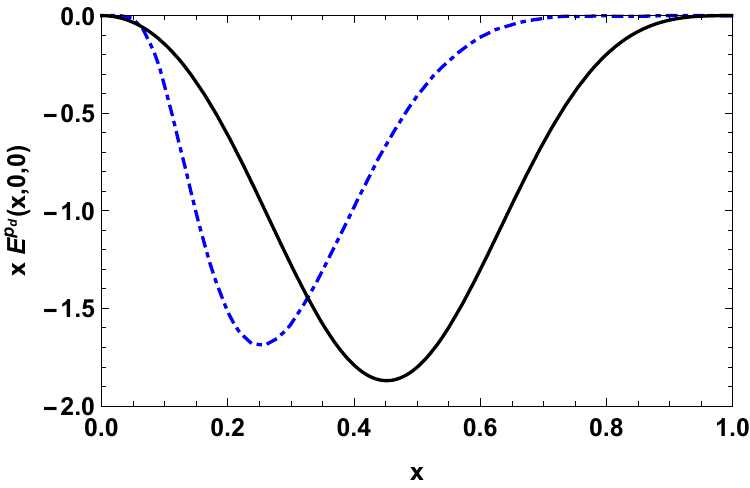}
		\hspace{0.03cm}		
	\end{minipage}
	\caption{\label{fig9ComGpd} (Color online) Comparison of our results (Solid black line) including contribution of timelike polarized diquarks in spin non-flip (Row $1$) and spin flip (Row $2$) generalized parton distributions with results of Ref. \cite{Lu:2010dt} (Dashed blue line) without involvement of timelike polarized diquarks.}
\end{figure*} 

Fig. \ref{fig8Essu} illustrates the effect of skewness and momentum transfer on the spin-flip GPDs of valence quarks including the contributions of only vector diquarks. The $d$ quark flavor in $p$, $s$ quark flavor in $\Sigma$ and $u$ quark flavor in $\Xi$ follow a similar trend as in Fig. \ref{fig5GPDsHssu} while carrying a share of longitudinal momentum fraction from their diquark systems. With the increase of skewness, $\zeta=0.01$ to $0.1$, an increase in the amplitude and peak value of $x$ are also observed in moving from left to right panel for both the cases of momentum transfer $\Dp=0.3$ and $0.6$. A fast decrement with an increase in $\Dp$ from $0.3$ to $0.6$ is also found for $p$ as compared to baryons carrying $s$ quark(s) while looking from top to bottom for two fixed values of skewness $\zeta=0.01$ and $0.1$, presented in left and right panels respectively. In a nut shell, with an increase in skewness, the amplitudes of both spin non-flip and flip GPDs get amplified. However, this amplification is found to be more in the spin non-flip GPDs due to the presence of a factor $\sqrt{1-\zeta}$ in Eq. (\ref{GPD_H}). 

Fig. \ref{fig9ComGpd} represents the comparison of our calculation corresponding to both GPDs with the results of Ref. \cite{Lu:2010dt} (without including timelike polarized diquarks) for the constituent $u$ and $d$ quark flavors of proton in columns $1$ and $2$ respectively. On comparing the results, we find that not much difference arises in the amplitudes of the distributions for $u$ quark flavor that include contributions of both scalar and vector diquark, but peaks are shifted towards higher values of $x$. Whereas in the case of GPD corresponding to $d$ quark flavor that include vector diquark only, a significant shift in the value of amplitude as well as capability of carrying $x$ has been observed.

\section{Longitudinal boost\label{SecLong_space}} 
By taking the FT with respect to $\zeta$, one can analyze the DVCS amplitudes in the longitudinal position space. The measure of longitudinal momentum transfer can be defined by the boost invariant longitudinal impact parameter, $\sigma=b^-P^+/2$ \cite{Manohar:2010zm}. An interesting outcome in the form of diffraction pattern of $\sigma$ parameter was first introduced in Ref. \cite{Brodsky:2006in}. The correlation determined in the three-dimensional $|\bfb|-\sigma$ space is frame independent as Lorentz boosts are kinematical in the front form. In longitudinal position space, spin non-flip and flip GPDs are expressed as \cite{Chakrabarti:2008mw}
\be 
H^{X_q}(x,\sigma,-t) &=& \frac{1}{2\pi} \int_{0}^{\zeta^\prime} d\zeta ~ e^{i\zeta P^+ b^-/2} H^{X_q}(x,\zeta,-t=\Dp^2) \, , \nonumber \\
&=& \frac{1}{2\pi} \int_{0}^{\zeta^\prime} d\zeta ~ e^{i\zeta \sigma} H(x,\zeta,-t=\Dp^2) \, , \\
E^{X_q}(x,\sigma,-t) &=& \frac{1}{2\pi} \int_{0}^{\zeta^\prime} d\zeta ~ e^{i\zeta P^+ b^-/2} E(x,\zeta,-t=\Dp^2) \, , \nonumber \\
&=& \frac{1}{2\pi} \int_{0}^{\zeta^\prime} d\zeta ~ e^{i\zeta \sigma} E^{X_q}(x,\zeta,-t=\Dp^2) \, ,
\ee  
where $\zeta^\prime$ is the upper limit of the integration. In optics, a finite slit width is an essential condition for the formation of a diffraction pattern. Here, the integration limit from $0$ to $\zeta^\prime$ will act as a required finite slit width for the formation of diffraction pattern corresponding to the FT DVCS amplitudes. Since, we are dealing in the domain $\zeta < x < 1$, the upper limit of integration will be $\zeta^\prime=\zeta_{max}$ if $x > \zeta_{max}$, otherwise upper limit of integration will be $\zeta^\prime=x$ if $x < \zeta_{max}$. For a fixed value of momentum transfer, $\zeta_{max}$ can be evaluated using
\be 
\zeta_{max}=\frac{-t}{2M^2_X} ~ \Bigg(\sqrt{1+\frac{4M_X^2}{(-t)}}-1\Bigg) \, .
\ee 
\begin{figure*}
	\centering
	\begin{minipage}[c]{0.98\textwidth}
		(a)\includegraphics[width=7.5cm]{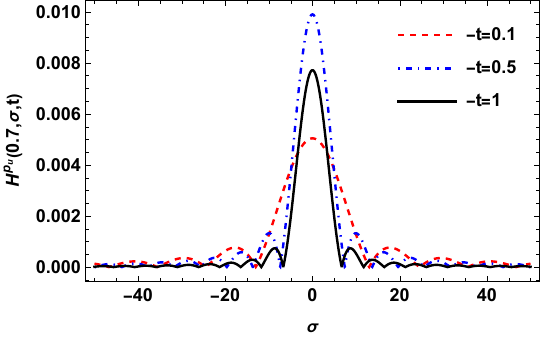}
		\hspace{0.03cm}
		(b)\includegraphics[width=7.5cm]{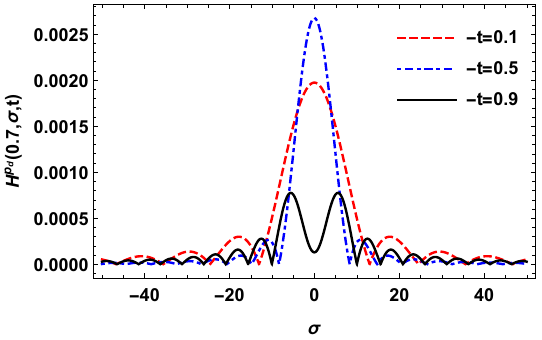}
		\hspace{0.03cm}
		(c)\includegraphics[width=7.5cm]{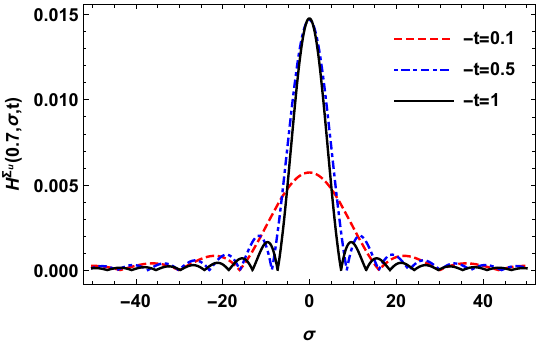}
		\hspace{0.03cm}
		(d)\includegraphics[width=7.5cm]{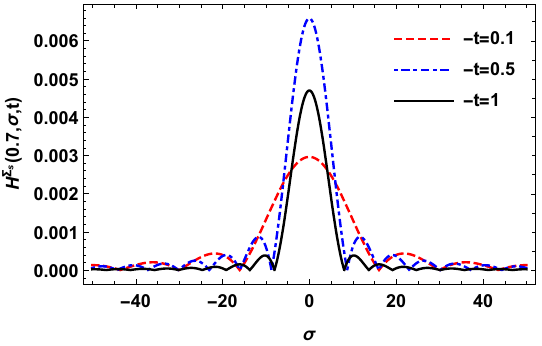}
		\hspace{0.03cm}	
		(e)\includegraphics[width=7.5cm]{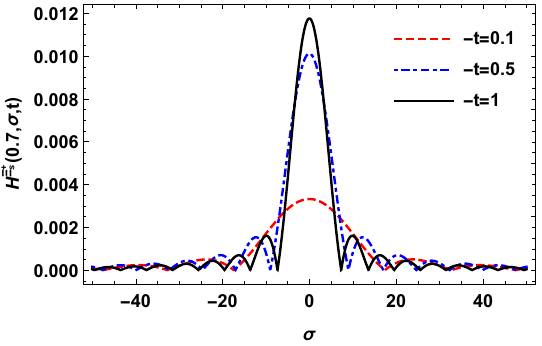}
		\hspace{0.03cm}
		(f)\includegraphics[width=7.5cm]{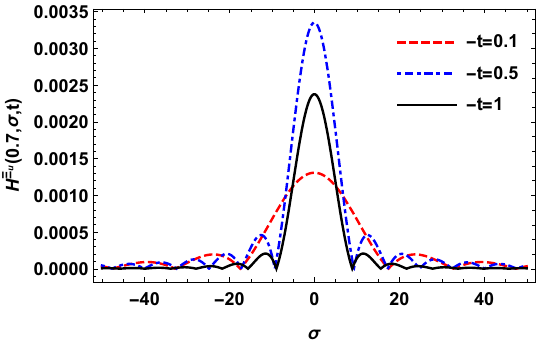}
		\hspace{0.03cm}				
	\end{minipage}
	\caption{\label{fig10LB_H} (Color online) Plots of chiral-even spin non-flip GPDs in longitudinal impact parameter space at fixed longitudinal momentum fraction $x=0.7$ for different values of $-t$ (GeV$^2$).}
\end{figure*} 
\begin{figure*}
	\centering
	\begin{minipage}[c]{0.98\textwidth}
		(a)\includegraphics[width=7.5cm]{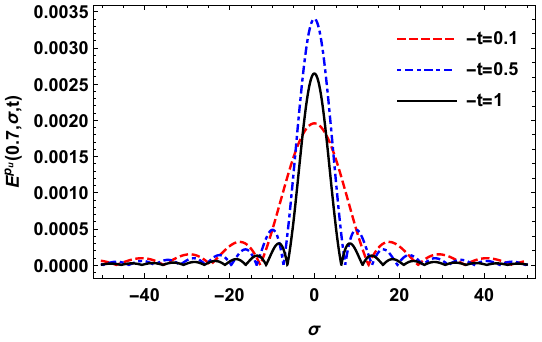}
		\hspace{0.03cm}
		(b)\includegraphics[width=7.5cm]{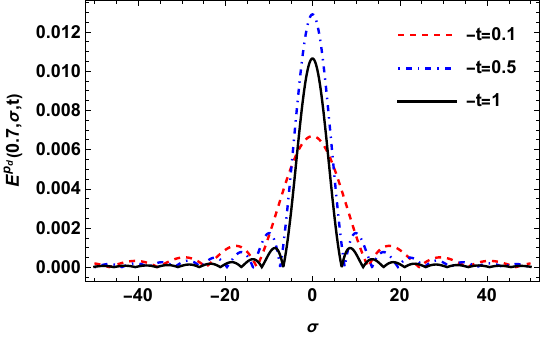}
		\hspace{0.03cm}
		(c)\includegraphics[width=7.5cm]{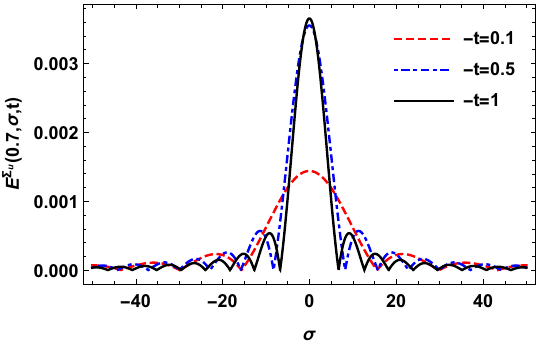}
		\hspace{0.03cm}	
		(d)\includegraphics[width=7.5cm]{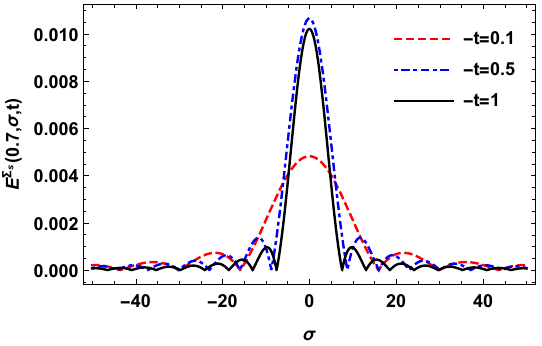}
		\hspace{0.03cm}
		(e)\includegraphics[width=7.5cm]{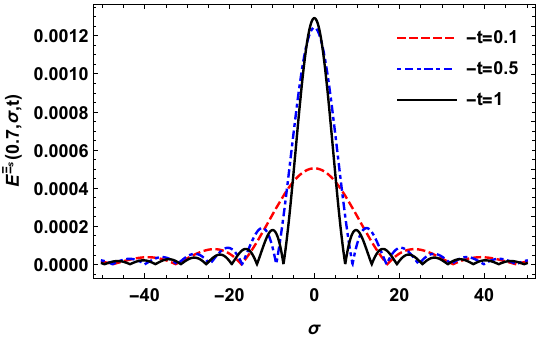}
		\hspace{0.03cm}
		(f)\includegraphics[width=7.5cm]{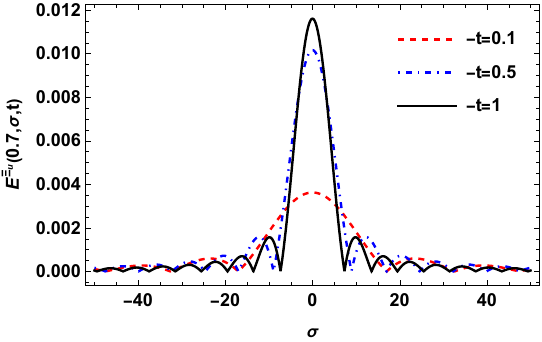}
		\hspace{0.03cm}			
	\end{minipage}
	\caption{\label{fig11LBE} (Color online) Plots of chiral-even spin flip GPDs in longitudinal impact parameter space at fixed longitudinal momentum fraction $x=0.7$ for different values of $-t$ (GeV$^2$).}
\end{figure*} 
For the case of $p$, the magnitude of GPDs in longitudinal position space corresponding to three different values of $-t$ (in GeV$^2$)has been demonstrated in Fig. \ref {fig10LB_H} (a) and \ref {fig10LB_H} (b) for $u$ and $d$ quark flavor respectively. As expected, all the distributions have a primary maxima at $\sigma=0$ for the $u$ quark flavor. With an increase in the value of $-t$, the distributions narrow down with drifting of other maxima towards $\sigma=0$. For the $d$ quark flavor, at smaller values of $-t$, the behavior of distribution is similar to the  $u$ quark flavor, however, with a smaller amplitude. At $-t=0.9$ GeV$^2$, a double peak structure has been observed, but at different value of $-t$ as analyzed in Ref. \cite{Mondal:2015uha}. Over all, the amplitude of the distributions is found to be less by a factor of 10 than as expressed in Ref. \cite{Mondal:2015uha} which may be attributed to different model scales and baselines. The distributions of spin-flip GPDs in the longitudinal impact parameter space for constituent quark flavors of $p$ are presented in Fig. \ref {fig11LBE} (a) and \ref {fig11LBE} (b). The amplitude for $d$ quark flavor is found to be more than $u$ quark flavor, otherwise the characteristics of the distributions are same for both flavors.

For strange baryons, the distributions corresponding to the magnitude of spin non-flip GPDs in longitudinal impact parameter space have been demonstrated in Figs. \ref{fig10LB_H} (c)-\ref{fig10LB_H} (f). The general trend of narrowing down of distributions is same as observed for the case of $p$ with an increase in the value of $-t$. However, the amplitude of distributions is found to be more and an increment in the value of amplitude has also been observed till $-t=1$ GeV$^2$. Double peak structure for the active quark flavors carrying contribution of only vector diquark is also observed to be missing in the region of $0.1<-t<1$. The distributions of spin flip GPDs in the longitudinal impact parameter space for constituent quark flavors of $\Sigma$ and $\Xi$ are presented in Fig. \ref {fig11LBE} (c)-\ref {fig11LBE} (f). Except for the amplitudes,  the distributions are qualitatively found to be consistent.

\section{Summary \label{SecSumm}}
This work presents a comparative analysis of electromagnetic form factors (EMFFs) and generalized parton distributions (GPDs) among low-lying octet baryons. The analysis has been performed by employing a diquark spectator model in the framework of light-cone dynamics. Dirac and Pauli form factor values (FFs) at zero momentum transfer have been computed by considering charge symmetry among isospin partners of octet baryons. The values of Dirac and Pauli FFs with respect to the square of momentum transfer are compared with the available data of constituent quark flavors of proton and strange baryons. The FFs are found to be aligned with the lattice simulations except for the $s$ quark flavor of $\Xi$ which may be attributed to the difference of our model assumption. Model calculation of the evolved unpolarized parton distribution function (PDF) for $u$ quark flavor of the proton has been compared with CTEQ5L parameterization to estimate the model scale. Comparison of PDFs corresponding to constituent quark flavors among octet baryons implies that heavier the quark, more longitudinal momentum fraction it will carry. GPDs with non-zero skewness are studied in the DGLAP region. With an increase in the value of the skewness parameter as well as the transverse momentum transfer, peak values of distributions are found to shift on higher values of longitudinal momentum fraction. However, a decrease in the amplitude has been found if transverse momentum transfer is increased with fixed skewness. An enhancement in the amplitude has been found if skewness is increased with fixed transverse momentum transfer. The characteristic of carrying a high longitudinal momentum fraction by a heavier quark is also found in the plots of GPDs. The GPDs in the longitudinal impact parameter space show a diffraction pattern with primary maxima at zero $\sigma$ for both spin flip and spin non-flip GPDs except for $d$ quark flavor of proton. The distributions corresponding to a proton are found to fall more quickly with an increase of $-t$ than strange baryons. No double peak structure has been observed for the strange baryons below $-t=1$ GeV$^2$. Experimental access regarding the measurement of distribution functions for strange baryons is minimal because of their short life times. Due to the lack of availability of EMFF measurement regarding strange baryons, we compare our results with available lattice data which exhibit favorable consistency. Furthermore, exploring additional strange baryon properties within the theoretical framework presents an intriguing avenue for further investigation.

\section{Acknowledgement}
H.D. would like to thank  the Science and Engineering Research Board, Anusandhan-National Research Foundation, Government of India under the scheme SERB-POWER Fellowship (Ref No. SPF/2023/000116) for financial support.

\section*{Appendix: Polarization vectors  \label{secApp}}
The four-vector light-cone polarization vectors $\varepsilon^\mu (k,\lambda_\mathfrak{a})$ used in the deduction of light-cone wave functions are given as follow 
\be 
\varepsilon^\mu (P-k,+) &=& \biggl[0, \frac{k_1 + i k_2}{\sqrt{2}(1-x)P^+}, -\frac{1}{2}, -\frac{i}{2}\biggr] \, , \nonumber \\
\varepsilon^\mu (P-k,-) &=& \biggl[0, -\frac{k_1 - i k_2}{\sqrt{2}(1-x)P^+}, \frac{1}{2}, -\frac{i}{2}\biggr] \, , \nonumber \\
\varepsilon^\mu (P-k,0) &=& \frac{1}{M_X} \biggl[(1-x)P^+, \frac{\bfk^2-M_X^2}{2(1-x)P^+}, -p_1, -p_2\biggr] \,  .
\ee 
These polarization vectors satisfy $\varepsilon^\mu (k,\lambda_\mathfrak{a}) \cdot \varepsilon_\mu (k,\lambda^\prime_\mathfrak{a}) =-\delta_{\lambda_\mathfrak{a} \lambda^\prime_\mathfrak{a}}$ and $(P-k) \cdot \varepsilon^\mu (k,\lambda_\mathfrak{a})=0$.  For completeness relation of the polarization sum in relation $d^{\mu \nu}=-g^{\mu \nu}$, the forth unphysical timelike polarization state is defined as $\varepsilon^\mu (P-k,t)=(P-k)^\mu /M_X$.



\begin{thebibliography}{200}
\section*{References}

\bibitem{Lorce:2013pza}
C.~Lorc\'e and B.~Pasquini,
JHEP \textbf{09}, 138 (2013).

\bibitem{Echevarria:2016mrc}
M.~G.~Echevarria, A.~Idilbi, K.~Kanazawa, C.~Lorc\'e, A.~Metz, B.~Pasquini and M.~Schlegel,
Phys. Lett. B \textbf{759}, 336-341 (2016).

\bibitem{Meissner:2009ww}
S.~Meissner, A.~Metz and M.~Schlegel,
JHEP \textbf{08}, 056 (2009).

\bibitem{Collins:2007ph}
J.~C.~Collins, T.~C.~Rogers and A.~M.~Stasto,
Phys. Rev. D \textbf{77}, 085009 (2008).

\bibitem{Collins:1981uw}
J.~C.~Collins and D.~E.~Soper,
Nucl. Phys. B \textbf{194}, 445-492 (1982).


\bibitem{Bacchetta:2024fci}
A.~Bacchetta, F.~G.~Celiberto and M.~Radici,
Eur. Phys. J. C \textbf{84}, 576 (2024).

\bibitem{Cerutti:2022lmb}
M.~Cerutti \textit{et al.} [MAP (Multi-dimensional Analyses of Partonic distributions)],
Phys. Rev. D \textbf{107}, 014014 (2023).

\bibitem{Ji:2004wu}
X.~d.~Ji, J.~p.~Ma and F.~Yuan,
Phys. Rev. D \textbf{71}, 034005 (2005).

\bibitem{HERMES:2003gbu}
A.~Airapetian \textit{et al.} [HERMES],
Phys. Rev. Lett. \textbf{92}, 012005 (2004).

\bibitem{SpinMuon:1997yns}
B.~Adeva \textit{et al.} [Spin Muon],
Phys. Lett. B \textbf{420}, 180-190 (1998).


\bibitem{Puhan:2023ekt}
S.~Puhan, S.~Sharma, N.~Kaur, N.~Kumar and H.~Dahiya,
JHEP \textbf{02}, 075 (2024).

\bibitem{Sharma:2023wha}
S.~Sharma, N.~Kumar and H.~Dahiya,
Nucl. Phys. B \textbf{992}, 116247 (2023).

\bibitem{Sharma:2022ylk}
S.~Sharma and H.~Dahiya,
Int. J. Mod. Phys. A \textbf{37}, 2250205 (2022).

\bibitem{Bacchetta:2019sam}
A.~Bacchetta, V.~Bertone, C.~Bissolotti, G.~Bozzi, F.~Delcarro, F.~Piacenza and M.~Radici,
JHEP \textbf{07}, 117 (2020).

\bibitem{Ji:1998pc}
X.~D.~Ji,
J. Phys. G \textbf{24}, 1181-1205 (1998).

\bibitem{Braunschweig:1985nr}
T.~Braunschweig, B.~Geyer, J.~Horejsi and D.~Robaschik,
Z. Phys. C \textbf{33}, 275 (1986).

\bibitem{Collins:1996fb}
J.~C.~Collins, L.~Frankfurt and M.~Strikman,
Phys. Rev. D \textbf{56}, 2982-3006 (1997).

\bibitem{Radyushkin:1996ru}
A.~V.~Radyushkin,
Phys. Lett. B \textbf{385}, 333-342 (1996).

\bibitem{Blumlein:1999sc}
J.~Blumlein, B.~Geyer and D.~Robaschik,
Nucl. Phys. B \textbf{560}, 283-344 (1999).

\bibitem{Hessberger:2016ypd}
F.~P.~He\ss{}berger, S.~Antalic, A.~K.~Mistry, D.~Ackermann, B.~Andel, M.~Block, Z.~Kalaninova, B.~Kindler, I.~Kojouharov and M.~Laatiaoui, \textit{et al.}
Eur. Phys. J. A \textbf{52}, 192 (2016).

\bibitem{H1:2007vrx}
F.~D.~Aaron \textit{et al.} [H1],
Phys. Lett. B \textbf{659}, 796-806 (2008).

\bibitem{JeffersonLabHallA:2006prd}
C.~Mu\~noz Camacho \textit{et al.} [Jefferson Lab Hall A and Hall A DVCS],
Phys. Rev. Lett. \textbf{97}, 262002 (2006).

\bibitem{Collins:1998be}
J.~C.~Collins and A.~Freund,
Phys. Rev. D \textbf{59}, 074009 (1999).

\bibitem{Goloskokov:2007nt}
S.~V.~Goloskokov and P.~Kroll,
Eur. Phys. J. C \textbf{53}, 367-384 (2008).

\bibitem{Vanderhaeghen:1999xj}
M.~Vanderhaeghen, P.~A.~M.~Guichon and M.~Guidal,
Phys. Rev. D \textbf{60}, 094017 (1999).

\bibitem{H1:2001nez}
C.~Adloff \textit{et al.} [H1],
Phys. Lett. B \textbf{517}, 47-58 (2001).

\bibitem{ZEUS:2003pwh}
S.~Chekanov \textit{et al.} [ZEUS],
Phys. Lett. B \textbf{573}, 46-62 (2003).

\bibitem{HERMES:2012gbh}
A.~Airapetian \textit{et al.} [HERMES],
JHEP \textbf{07}, 032 (2012).

\bibitem{HERMES:2011bou}
A.~Airapetian \textit{et al.} [HERMES],
Phys. Lett. B \textbf{704}, 15-23 (2011).

\bibitem{Diehl:2003ny}
M.~Diehl,
Phys. Rept. \textbf{388}, 41-277 (2003).


\bibitem{Mamo:2022jhp}
K.~A.~Mamo and I.~Zahed,
Phys. Rev. D \textbf{108}, 086026 (2023).

\bibitem{Kumericki:2007sa}
K.~Kumericki, D.~Mueller and K.~Passek-Kumericki,
Nucl. Phys. B \textbf{794}, 244-323 (2008).

\bibitem{Mueller:2005ed}
D.~Mueller and A.~Schafer,
Nucl. Phys. B \textbf{739}, 1-59 (2006).

\bibitem{Freund:2001bf}
A.~Freund and M.~F.~McDermott,
Phys. Rev. D \textbf{65}, 056012 (2002)
[erratum: Phys. Rev. D \textbf{66}, 079903 (2002)].

\bibitem{Duplancic:2023xrt}
G.~Duplan\v{c}i\'c, P.~Kroll, K.~Passek-K. and L.~Szymanowski,
Phys. Rev. D \textbf{109}, 034008 (2024).

\bibitem{Sharma:2023ibp}
S.~Sharma and H.~Dahiya,
Nucl. Phys. B \textbf{1001}, 116522 (2024).

\bibitem{Pire:2013vea}
B.~Pire, L.~Szymanowski and S.~Wallon,
[arXiv:1309.0083 [hep-ph]].

\bibitem{Kaur:2018ewq}
N.~Kaur, N.~Kumar, C.~Mondal and H.~Dahiya,
Nucl. Phys. B \textbf{934}, 80-95 (2018).

\bibitem{Mondal:2017wbf}
C.~Mondal,
Eur. Phys. J. C \textbf{77}, 640 (2017).

\bibitem{Kumar:2015yta}
N.~Kumar and H.~Dahiya,
Phys. Rev. D \textbf{91}, 114031 (2015).

\bibitem{Manohar:2010zm}
R.~Manohar, A.~Mukherjee and D.~Chakrabarti,
Phys. Rev. D \textbf{83}, 014004 (2011).

\bibitem{Chakrabarti:2008mw}
D.~Chakrabarti, R.~Manohar and A.~Mukherjee,
Phys. Rev. D \textbf{79}, 034006 (2009).

\bibitem{Brodsky:2006ku}
S.~J.~Brodsky, D.~Chakrabarti, A.~Harindranath, A.~Mukherjee and J.~P.~Vary,
Phys. Rev. D \textbf{75}, 014003 (2007).

\bibitem{Brodsky:2006in}
S.~J.~Brodsky, D.~Chakrabarti, A.~Harindranath, A.~Mukherjee and J.~P.~Vary,
Phys. Lett. B \textbf{641}, 440-446 (2006).

\bibitem{Barry:2023qqh}
P.~C.~Barry \textit{et al.} [Jefferson Lab Angular Momentum (JAM)],
Phys. Rev. D \textbf{108}, L091504 (2023).

\bibitem{Bor:2022fga}
J.~Bor and D.~Boer,
Phys. Rev. D \textbf{106}, 014030 (2022).

\bibitem{GlueX:2019mkq}
A.~Ali \textit{et al.} [GlueX],
Phys. Rev. Lett. \textbf{123}, 072001 (2019).

\bibitem{Moutarde:2019tqa}
H.~Moutarde, P.~Sznajder and J.~Wagner,
Eur. Phys. J. C \textbf{79}, 614 (2019).


\bibitem{HERMES:2001bob}
A.~Airapetian \textit{et al.} [HERMES],
Phys. Rev. Lett. \textbf{87}, 182001 (2001).

\bibitem{CSSM:2014knt}
P.~E.~Shanahan \textit{et al.} [CSSM and QCDSF/UKQCD],
Phys. Rev. D \textbf{89}, 074511 (2014).

\bibitem{Shanahan:2014cga}
P.~E.~Shanahan, A.~W.~Thomas, R.~D.~Young, J.~M.~Zanotti, R.~Horsley, Y.~Nakamura, D.~Pleiter, P.~E.~L.~Rakow, G.~Schierholz and H.~St\"uben,
Phys. Rev. D \textbf{90}, 034502 (2014).

\bibitem{Han:2024ucv}
C.~Han, W.~Wang, J.~Zeng and J.~L.~Zhang,
JHEP \textbf{07}, 019 (2024).

\bibitem{Zhu:2023nhl}
Z.~Zhu \textit{et al.} [BLFQ],
Phys. Rev. D \textbf{108}, 036009 (2023).

\bibitem{Carrillo-Serrano:2016igi}
M.~E.~Carrillo-Serrano, W.~Bentz, I.~C.~Clo\"et and A.~W.~Thomas,
Phys. Lett. B \textbf{759}, 178-183 (2016).

\bibitem{Zhang:2016qqg}
J.~Zhang and B.~Q.~Ma,
Phys. Rev. C \textbf{93}, 065209 (2016).

\bibitem{Jiang:2009jn}
F.~J.~Jiang and B.~C.~Tiburzi,
Phys. Rev. D \textbf{81}, 034017 (2010).

\bibitem{Feliciello:2015dua}
A.~Feliciello and T.~Nagae,
Rept. Prog. Phys. \textbf{78}, 096301 (2015).

\bibitem{Wang:2005vg}
P.~Wang, S.~Lawley, D.~B.~Leinweber, A.~W.~Thomas and A.~G.~Williams,
Phys. Rev. C \textbf{72}, 045801 (2005).

\bibitem{Dirac:1949cp}
P.~A.~M.~Dirac,
Rev. Mod. Phys. \textbf{21}, 392-399 (1949).

\bibitem{Bacchetta:2008af}
A.~Bacchetta, F.~Conti and M.~Radici,
Phys. Rev. D \textbf{78}, 074010 (2008).

\bibitem{Jakob:1997wg}
R.~Jakob, P.~J.~Mulders and J.~Rodrigues,
Nucl. Phys. A \textbf{626}, 937-965 (1997).

\bibitem{Bacchetta:2003rz}
A.~Bacchetta, A.~Schaefer and J.~J.~Yang,
Phys. Lett. B \textbf{578}, 109-118 (2004).

\bibitem{Brodsky:2000ii}
S.~J.~Brodsky, D.~S.~Hwang, B.~Q.~Ma and I.~Schmidt,
Nucl. Phys. B \textbf{593}, 311-335 (2001).

\bibitem{Lichtenberg:1968zz}
D.~B.~Lichtenberg, L.~J.~Tassie and P.~J.~Keleman,
Phys. Rev. \textbf{167}, 1535-1542 (1968).








\bibitem{Cates:2011pz}
G.~D.~Cates, C.~W.~de Jager, S.~Riordan and B.~Wojtsekhowski,
Phys. Rev. Lett. \textbf{106}, 252003 (2011).





\bibitem{Brodsky:1980zm}
S.~J.~Brodsky and S.~D.~Drell,
Phys. Rev. D \textbf{22}, 2236 (1980).




\bibitem{Dahiya:2007is}
H.~Dahiya, A.~Mukherjee and S.~Ray,
Phys. Rev. D \textbf{76}, 034010 (2007).

\bibitem{Hwang:2007tb}
D.~S.~Hwang and D.~Mueller,
Phys. Lett. B \textbf{660}, 350-359 (2008).

\bibitem{Lai:1999wy}
H.~L.~Lai \textit{et al.} [CTEQ],
Eur. Phys. J. C \textbf{12}, 375-392 (2000).

\bibitem{Mondal:2015uha}
C.~Mondal and D.~Chakrabarti,
Eur. Phys. J. C \textbf{75}, 261 (2015).

\bibitem{Lu:2010dt}
Z.~Lu and I.~Schmidt,
Phys. Rev. D \textbf{82}, 094005 (2010).


\end{thebibliography}
\end{document}